\documentclass[lettersize,journal]{IEEEtran}
\usepackage{amsmath,amsfonts}
\usepackage{algorithmic}
\usepackage{algorithm}
\usepackage{array}
\usepackage{textcomp}
\usepackage{stfloats}
\usepackage{url}
\usepackage{verbatim}
\usepackage{graphicx}
\usepackage{cite}
\usepackage{amsthm,amsmath,amssymb}
\usepackage{mathrsfs}
\usepackage{enumitem}
\usepackage{subcaption} 
\usepackage{times}
\usepackage{svg}
\usepackage{multirow}
\usepackage{amsthm}            
\usepackage{longtable}
\usepackage{booktabs}

\theoremstyle{plain}

\newtheorem{definition}{Definition}

\newtheorem{proposition}{Proposition}
\begin{document}

\title{Green Emergency Communications in RIS- and MA-Assisted Multi-UAV SAGINs: A Partially Observable Reinforcement Learning Approach}

\author{Liangshun Wu, Wen Chen, Shunqing Zhang, Yajun Wang, Kunlun Wang\\
%
%
\thanks{Liangshun Wu and Wen Chen are with School of Information and Electronic Engineering, Shanghai Jiao Tong University, Shanghai, China. Email: wuliangshun@sjtu.edu.cn, wenchen@sjtu.edu.cn. Shunqing Zhang is with School of Communication and Information Engineering, Shanghai University, Shanghai, China. Email: shunqing@shu.edu.cn. Yajun Wang is with Jiangsu University of Science and Technology, Zhenjiang, China. Email:wangyj1859@just.edu.cn. Kunlun Wang is with School of Communication and Electronic Engineering, East China Normal University, Shanghai, China. Email: klwang@cee.ecnu.edu.cn.}
%
}


\maketitle

\begin{abstract}
In post-disaster space–air–ground integrated networks (SAGINs), terrestrial infrastructure is often impaired, and unmanned aerial vehicles (UAVs) must rapidly restore connectivity for mission-critical ground terminals in cluttered non-line-of-sight (NLoS) urban environments. To enhance coverage, UAVs employ movable antennas (MAs), while reconfigurable intelligent surfaces (RISs) on surviving high-rises redirect signals. The key challenge is communication-limited partial observability, leaving each UAV with a narrow, fast-changing neighborhood view that destabilizes value estimation. Existing multi-agent reinforcement learning (MARL) approaches are inadequate—non-communication methods rely on unavailable global critics, heuristic sharing is brittle and redundant, and learnable protocols (e.g., CommNet, DIAL) lose per-neighbor structure and aggravate non-stationarity under tight bandwidth. To address partial observability, we propose a spatiotemporal A2C where each UAV transmits prior-decision messages with local state, a compact policy fingerprint, and a recurrent belief, encoded per neighbor and concatenated. A spatial discount shapes value targets to emphasize local interactions, while analysis under one-hop-per-slot latency explains stable training with delayed views. Experimental results show our policy outperforms IA2C, ConseNet, FPrint, DIAL, and CommNet—achieving faster convergence, higher asymptotic reward, reduced Temporal-Difference(TD)/advantage errors, and a better communication throughput–energy trade-off.
\end{abstract}

\section{Introduction}

\begin{figure*}[h]
\centering
\includegraphics[width=\textwidth]{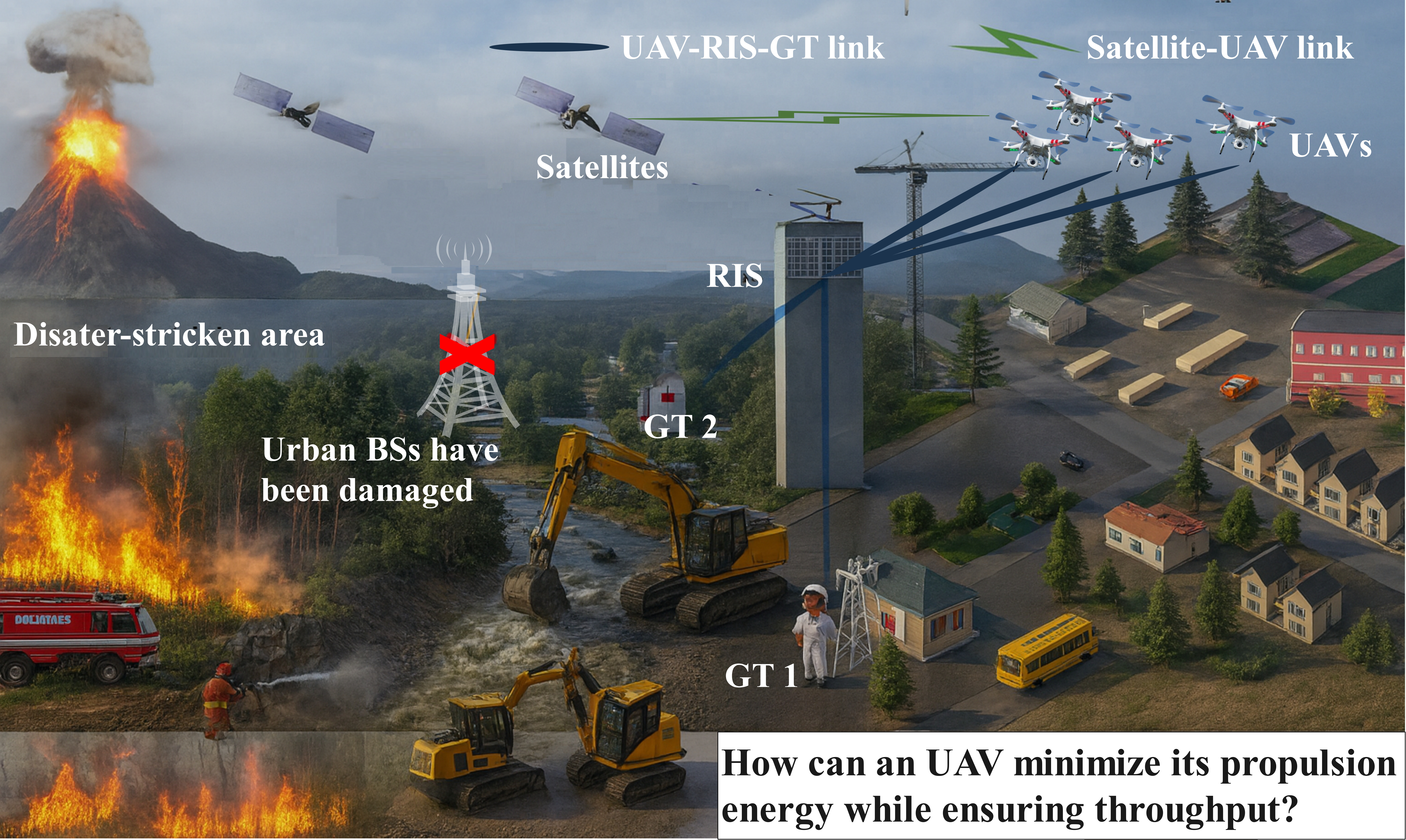} %
\caption{SAGIN emergency communication scenario.}\label{scene}
\end{figure*}

\IEEEPARstart{I}{n} the construction of modern smart cities, sudden disasters such as volcanic eruptions and earthquakes often severely damage urban communication infrastructure. Base stations (BSs) are highly vulnerable to power outages or tower collapses, leading to the rapid formation of large-scale communication blind zones. Meanwhile, mission-critical ground terminals (GTs)—including hospitals, fire departments, rescue teams, and unmanned rescue robots—require reliable, low-latency data links and computational resources to support time-sensitive tasks such as situational awareness, path planning, and damage assessment. However, in disaster-stricken areas, network congestion and latency surges render traditional cloud computing insufficient to meet urgent rescue demands. To overcome these challenges, emergency management agencies can rapidly deploy unmanned aerial vehicles (UAVs) equipped with mobile edge computing (MEC) capabilities, forming a temporary Space-Air-Ground Integrated Network (SAGIN)\cite{10964638}. In SAGIN as shown in Fig. \ref{scene}, satellites issue mission commands to UAVs, which in turn relay instructions to GTs and perform computation tasks; satellites remain uninvolved in communication optimization\cite{chen2025space}. With ongoing advances in UAV payload capacity and edge computing hardware, airborne MEC has emerged as a key enabler of emergency communications\cite{shi2025dynamic}.  UAV swarms can flexibly adapt to complex urban terrains and dynamic rescue environments, enabling rapid restoration of network coverage, reducing energy consumption and latency, and ensuring high-throughput emergency communication\cite{xia2024uav}, local computation\cite{shi2025dynamic}, and intelligent decision-making under dense user conditions\cite{betalo2025generative}—all of which are critical for timely disaster response.

Due to potential blockages by urban buildings and non-line-of-sight (NLoS) effects caused by UAVs flying at high altitudes, the UAV-GT link may suffer from degraded transmission rates\cite{mei20223d}. Reconfigurable Intelligent Surfaces (RIS) are  widely recognized for their low cost, ease of deployment, and passive reflecting units (PRUs) phased-array radar characteristics\cite{8811733}. Although disasters such as earthquakes and volcanic eruptions can cause large-scale collapses of high-rise buildings, it is unlikely that an entire city’s structures would be destroyed. Because high-rise buildings are generally designed with strong seismic resistance, many—particularly those outside the immediate disaster zone—are likely to remain standing. In such cases, RIS PRUs can be installed on structurally sound buildings near the core rescue area\cite{matracia2023comparing}. By dynamically adjusting their phase shifts to redirect signals, these PRUs can effectively mitigate the impact of ground obstacles and NLoS conditions, ensuring the reliable transmission of critical data and commands. Compared to mounting RIS on UAVs, deploying them on high-rise buildings provides broader ground coverage and lower costs, while avoiding endurance limitations faced by UAV-based platforms.

In traditional UAV relay communications, Fixed Position Antennas (FPAs) are commonly used; however, their limited directivity and channel gain impose constraints on achievable communication rates. Recent studies \cite{10278220,10906511} have introduced Movable Antennas (MAs) as a flexible alternative for enhancing wireless communication. Unlike FPAs, MAs can dynamically adjust their positions to increase channel gain and suppress interference, thereby improving communication performance \cite{10972180,10643473,11082461}. UAVs equipped with MAs are thus better adapted to the stochastic and time-varying conditions of post-disaster emergency communication environments.

\subsection{Partial Observability}
In the aforementioned SAGIN, UAVs act as temporary aerial BSs that provide data services to GTs. However, the presence of temporary obstacles and environmental noise at disaster sites, combined with limited spectrum resources and constrained network bandwidth, makes information synchronization among UAVs extremely difficult. As a result, global state sharing is often infeasible \cite{10925529,wang2023joint}.  

\subsection{MARL Addressing Partial Observability}
Reinforcement Learning (RL) has been widely applied for adaptive decision-making in dynamic environments. Nevertheless, single-agent RL does not scale well to multi-agent systems. Multi-Agent Reinforcement Learning (MARL) extends RL to decentralized settings by enabling agents to learn policies based on local observations. However, MARL itself faces fundamental challenges, including partial observability and non-stationarity, which significantly complicate both training and deployment. To address these issues, existing MARL communication mechanisms can be broadly classified into four categories:  
\begin{itemize}
    \item {Non-communication methods}: Rely on centralized critics (e.g., MADDPG \cite{lowe2017multi}, IA2C\cite{lowe2017multi}) and improved value estimation (e.g., COMA \cite{foerster2016learning}), or use shared global parameters (e.g., Dec-HDRQN \cite{epochsupplemental}, PS-TRPO \cite{gupta2017cooperative} and ConseNet\cite{zhang2018fully}). These approaches, however, require access to global information, which limits their applicability in realistic local-observation scenarios.
    \item {Direct/heuristic communication}: Employ predefined protocols or simple information sharing (e.g., policy fingerprints \cite{foerster2016learning} or mean-field approximations \cite{yang2018mean}). Such methods are generally inefficient and prone to redundant communication overhead.
    \item {Learnable communication protocols}: Enable agents to learn how to encode and exchange messages during training (e.g., DIAL \cite{foerster2016learning}, CommNet \cite{sukhbaatar2016learning}, BiCNet \cite{peng2017multiagent} and FPrint\cite{foerster2017stabilising}). However, these approaches risk information loss and instability due to aggregation or improper encoding strategies.
    \item {Attention-based communication}: Introduce attention mechanisms to prioritize communication partners (e.g., ATOC \cite{jiang2018learning}, IC3Net \cite{singh2018learning}). While effective in some cases, their performance degrades when communication is restricted to small local neighborhoods.
\end{itemize}

Although each category has advanced the field, all suffer from inherent limitations under highly dynamic, bandwidth-constrained, and latency-sensitive emergency communication environments. Moreover, safety and reliability concerns preclude online policy learning in disaster scenarios, necessitating offline training with subsequent deployment for real-time inference only.

\subsection{Contributions}
To address the challenge of partial observability in emergency communication scenarios, this work focuses on enabling UAV swarms to make efficient decisions based solely on local observations and limited neighbor information. Our objective is to ensure reliable communication throughput while minimizing UAV energy consumption, thereby achieving green and sustainable emergency communication.

Our main contributions are:
\begin{enumerate}
    \item We formalize the partially observable multi-UAV emergency communication problem as a decentralized networked Markov Decision Process (MDP) and incorporate a spatial discount factor, allowing each UAV to train stably and converge using locally available information, reducing sensitivity to distant UAVs’ delayed updates.
    \item We design a differentiable communication protocol that, unlike existing methods (e.g., CommNet\cite{sukhbaatar2016learning} and DIAL\cite{foerster2016learning}), allows each UAV to encode received neighbor messages individually without lossy pre-averaging, and to concatenate rather than sum encoded features. Crucially, we embed policy fingerprints into messages, which mitigates non-stationarity and avoids information loss — particularly important under bandwidth-limited, post-disaster conditions.
\end{enumerate}

\section{System Model and Problem Formulation}

\begin{table}[htbp]
\centering
\caption{List of symbols used}
\begin{tabular}{lp{5cm}}
\hline
{Symbol} & {Meaning}\\ \hline
$\mathcal{U}=\{1,\ldots,J\}$ & UAV indices\\
$\mathcal{K}=\{1,\ldots,K\}$, $k$ & GT set and index\\
$\mathcal{N}=\{1,\ldots,N\}$, $n,m,\tau$ & Time-slot set / indices\\
$\mathcal{L}=\{1,\ldots,L\}$ & Cell indices (AoI grid)\\
$\mathcal{Q}=\{\mathbf{p}_1,\ldots,\mathbf{p}_Q\}$, $Q$ & MA positions / count\\
$G(\mathcal{V},\mathcal{E})$ & UAV communication graph\\
$\mathcal{V}=\{1,\ldots,J\}$, $\mathcal{E}$ & Nodes; edges\\
$\mathcal{N}_j$, $\mathcal{V}_j$ & Neighbors of $j$; $\mathcal{N}_j\cup\{j\}$\\
$\operatorname{mode}(\cdot)$, $\operatorname{diag}(\cdot)$, $\otimes$, $\mathrm{H}(\cdot)$ & Mode, diag, Kronecker, Policy entropy\\
$\mathrm{anc}(\cdot)$ & Gradient ancestors\\
$\mathbf{L}_i^c=[x_i,y_i]^T$ & Center of cell $i$\\
$x_s=y_s=l_s$ & Cell spacings (x/y)\\ 
$\mathbf{L}_j^\mathrm{u}[n]$, $\mathbf{L}_0^\mathrm{u},\mathbf{L}_f^\mathrm{u}$ & horizontal position; initial/final positions\\
$h_0^{\mathrm{u}}$ & Initial altitude level for each UAV\\
$h_j^\mathrm{u}[n]$, $H_j^\mathrm{u}[n]$, $h_s$ & Altitude level/actual altitude/height per level\\
$t_j^\mathrm{u}[n]$, $t_{\min},t_{\max}$ & Slot duration; bounds\\
$V_{\max}^h,V_{\max}^v$ & Max horiz. / vert. speed\\
$v_j^{h}[n]$,$v_j^{v}[n]$, $e_j^\mathrm{u}[n]$ & Horizontal/Vertical speed/Propulsion energy\\
$P_0,P_1,P_2$ & Profile, induced, ascent/descent power\\
$U_{\text{tip}}$, $v_o$ & Blade tip speed; induced speed (hover)\\
$d_0, \rho, s, G$ & Drag ratio, air density, solidity, disc area\\
$M_c\times M_r$, $M_c,M_r$ & PRU layout; counts (cols/rows)\\
$d_c,d_r$ & PRU spacing (col/row)\\
$\mathbf{w}_R=[x_R,y_R]^T, z_R$ & RIS position (horiz., vert.)\\
$\overline{\boldsymbol{\Theta}}[n]=\operatorname{diag}(\overline{\boldsymbol{\omega}}[n])$ & Averaged RIS phase matrix\\
$\overline{\boldsymbol{\omega}}[n]=\frac{1}{J}\sum_{j}\boldsymbol{\omega}_j[n]$ & Avg. phase vector\\
$\boldsymbol{\omega}_j[n]$, $\omega_{m_r,m_c}^{(j)}[n]$ & UAV $j$ phase vector; PRU phase $[-\pi,\pi)$\\
$\mathbf{p}_q=[\Delta x_q,\Delta y_q]^T$ & MA offset option $q$\\
$a_{j,q}[n]\in\{0,1\}$ & MA position indicator; exclusivity\\
$\mathbf{A}_j[n]=\{a_{j,q}[n]\}$ & MA decision vector\\
$\mathbf{w}_k=[x_k,y_k]^T$, $D_k$ & GT $k$ position; data demand\\
$\widetilde{\mathbf{L}}_j^\mathrm{u}[n]$, $\widetilde{H}_j^\mathrm{u}[n]$ & Actual MA horiz./altitude\\
$d_j^\mathrm{ur}[n]$, $g_j^\mathrm{ur}[n]$ & UAV–RIS distance; array response\\
$\xi, D_0$ & Pathloss at ref. distance; reference distance\\
$[\phi_j^\mathrm{ur},\varphi_j^\mathrm{ur},\psi_j^\mathrm{ur}]^T$ & Direction vectors\\
$\vartheta, \theta, r$ & Azimuth/elevation angle, horiz. projection\\
$\lambda, \frac{2\pi}{\lambda}$ & wavelength, wavenumber\\
$\Delta \ell_r,\Delta \ell_c$ & Extra path (row/col) in incoming wave\\
$\Delta \varphi_r,\Delta \varphi_c$ & Phase step (row/col)\\
$g_k^\mathrm{rg}$, $d_k^\mathrm{rg}$ & RIS–GT array response; distance\\
$\phi_k^\mathrm{rg},\varphi_k^\mathrm{rg},\psi_k^\mathrm{rg}$ & RIS–GT AoD direction cosines (x,y,z)\\
$g_{j,k}^\mathrm{urg}[n]$ & UAV–RIS–GT cascaded gain\\
$a$ (cascade) & RIS reflection amplitude/scaling\\
$p_{j,k}[n]$ & LoS blockage probability (UAV $j$ to GT $k$)\\
$d_{j,k}^\mathrm{ug}$ & UAV–GT distance\\
$a,b$ (blockage) & Environment constants (LoS model)\\
$g_k[n]$ & Avg. effective gain at GT $k$\\
$r_k[n]$ & Rate of GT $k$ (slot $n$)\\
$P,B,\sigma^2$ & Tx power, bandwidth, noise variance\\
$c_{j,k}[n],\, c_k[n]$ & Vote of UAV $j$; final schedule (TDMA)\\
$a_j^{\text{hor}}[n]$, $a_j^{\text{ver}}[n]$ & Horizontal / vertical action\\
$\mathbf{C}_j[n]=\{c_{j,k}[n]\}$ & UAV $j$ scheduling votes\\
$\boldsymbol{\omega}_j[n]$ & RIS phase recommendation\\
$\mathcal{S}_j[n]$, $\tilde{\mathcal{S}}_j[n]$ & Local state; augmented with neighbors\\
$\mathcal{A}_j[n]$ & Joint action (traj/MA/sched/time/RIS)\\
$\mathcal{M}_{ij}$, $\mathcal{M}_j[n]$ & Message $i\!\to\!j$; broadcast of $j$\\
$p_j(\cdot)$ & Local transition kernel\\
$\mathcal{R}_j[n]$, $\mathcal{R}[n]$ & Local reward; global average\\
$R$, $h$, $d_{ji}$ & Obs. radius; hop radius; graph distance\\
$\hat{\mathcal{R}}_j[n]$ & Spatiotemporally discounted reward\\
$\alpha,\gamma$ & Spatial / temporal discount\\
$v_j[\cdot]$, $\mathcal{B}$, $n_{\mathcal{B}}$ & Bootstrap value; minibatch; tail index\\
$\pi_{\theta_j}$, $V_{\phi_j}$ & Policy (actor); value (critic)\\
$\hat{A}_j[n]$, $\beta$ & Advantage; entropy weight\\
$\mathcal{L}(\theta_j)$, $\mathcal{L}(\phi_j)$ & Actor / critic losses\\
$h_j[n]$ & Prior-decision belief of UAV $j$\\
$e^{s}_{\lambda_j},e^{p}_{\lambda_j},e^{h}_{\lambda_j}$ & Encoders (state/policy/belief)\\
$g_{\nu_j}(\cdot)$, $\mu^{\mathrm{send}}_j,\mu^{\mathrm{recv}}_j$ & Aggregator; attentions\\
$K$, $h_j^{(k)}[n]$, $\nu^{(k)}_j$ & In-slot passes; refined belief; params\\
\hline
\end{tabular}
\label{tab:sym_marl}
\end{table}
\subsection{System Model}

\subsubsection{UAV Trajectory}
\begin{figure}[h]
\centering
\includegraphics[width=0.5\textwidth]{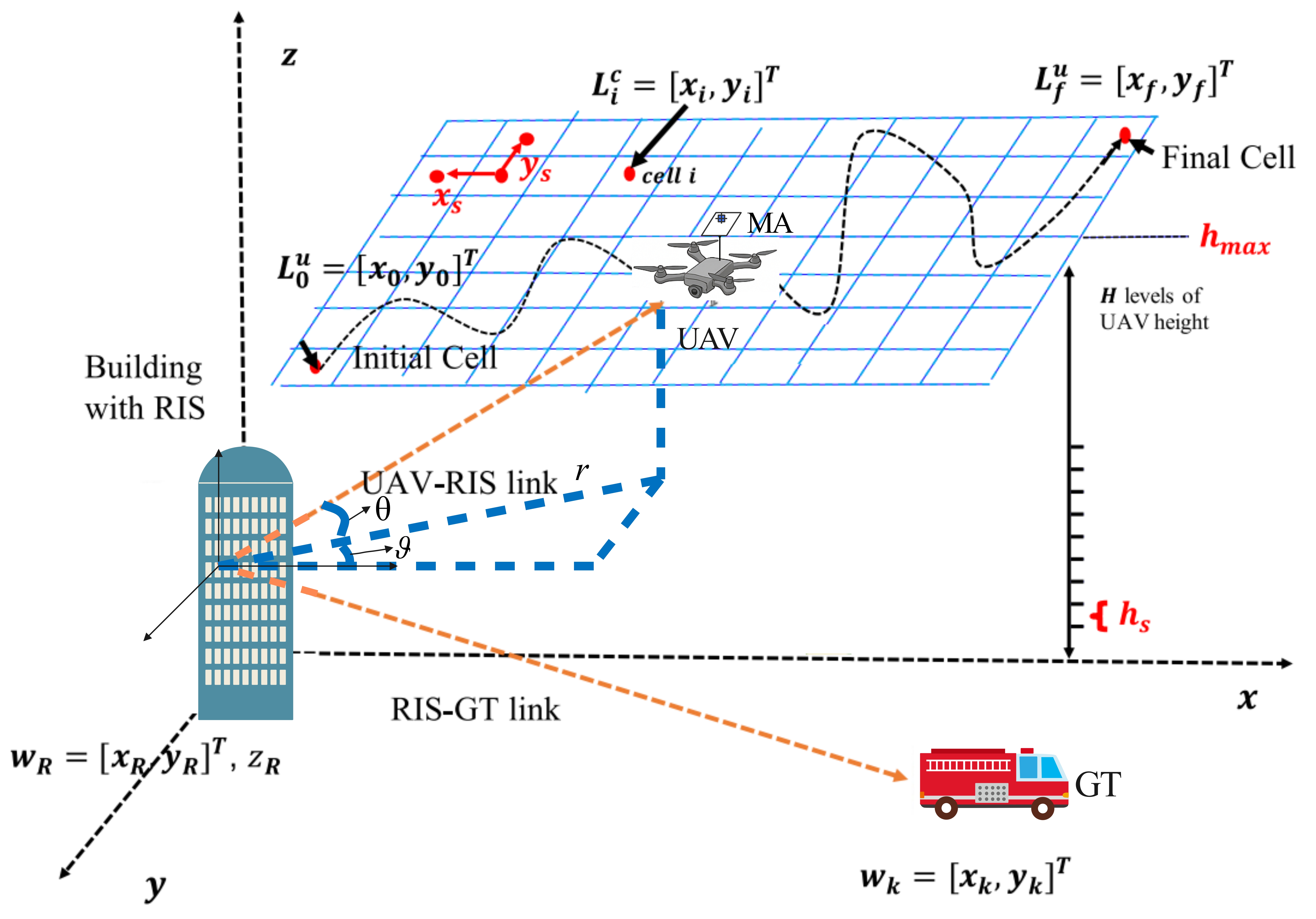} %
\caption{System model.}\label{fig:model}
\end{figure}

This paper considers the system shown in Fig. \ref{scene}. In this system, there are $J$ UAVs deployed to provide service to ground terminals (GTs). Let $\mathcal{U} = \{1, \ldots, J\}$ index the $J$ UAVs. Assume that the area of interest (AoI) is discretized into $L$ equal-sized cells (see Fig. \ref{fig:model}). Let $\mathbf{L}_i^c = [x_i, y_i]^T \in \mathbb{R}^{2 \times 1}$ represent the coordinates of the center of cell $i$; $x_s$ and $y_s$ are the distances between adjacent cells along the x and y axes, respectively. The horizontal position of UAV $j \in \mathcal{U}$ at time slot $n$ can be represented as $\mathbf{L}_j^\mathrm{u}[n] \in \mathcal{L}$, where $\mathcal{L} \triangleq \{1, 2, \ldots, L\}$ and $n \in \mathcal{N} \triangleq \{1, \ldots, N\}$, with $N$ being the total number of time slots. Note that $\mathbf{L}_0^\mathrm{u}$ and $\mathbf{L}_f^\mathrm{u}$ represent the pre-determined initial and final positions of all UAVs, respectively. In the vertical dimension, let $h_j^\mathrm{u}[n] \in \mathcal{H} \triangleq \{1, 2, \ldots, H\}$ be the height level of UAV $j$ at time slot $n$, where $H$ is the total number of height levels. Then, the height of UAV $j$ can be expressed as $H_j^\mathrm{u}[n] = h_j^\mathrm{u}[n] \cdot h_s$, where $h_s = \left\lfloor \frac{h_{\max}}{H} \right\rfloor$ is the vertical distance per height level. Additionally, let $t_j^\mathrm{u}[n]$ be the duration of time slot $n$, and assume it is sufficiently small, such that $t_{\min} \leq t_j^\mathrm{u}[n] \leq t_{\max}$. Thus, the total task completion time of UAV $j$ can be written as $\tau = \sum_{n=1}^{N} t_j^\mathrm{u}[n]$. Therefore, the trajectory of UAV $j$ can be represented by $N$ three-dimensional coordinates $\left[\mathbf{L}_j^\mathrm{u}[n], H_j^\mathrm{u}[n]\right]$, for all $n \in \mathcal{N}$, and the duration $t_j^\mathrm{u}[n]$ for each time slot.

Given the maximum horizontal speed $V_{\max}^h$ of UAV $j$, we can choose a sufficiently large number of time slots $N$, so that the position change of UAV $j$ within each time slot $t_j^\mathrm{u}[n]$ can be neglected with respect to the link distance between UAV $j$ and the GTs.

Based on the above assumptions, the horizontal flight speed of UAV $j$ at time slot $n$ is given by

\begin{equation}
v_j^h[n] = \frac{\left\| \mathbf{L}_j^\mathrm{u}[n+1] - \mathbf{L}_j^\mathrm{u}[n] \right\|}{t_j^\mathrm{u}[n]} \leq V_{\max}^h, \quad \forall n \in \mathcal{N}.
\end{equation}
If $v_j^h[n] = 0$, UAV $j$ will hover at time slot $n$. The vertical flight speed of UAV $j$ at time slot $n$ is

\begin{equation}
v_j^v[n] = \frac{\left\| H_j^\mathrm{u}[n+1] - H_j^\mathrm{u}[n] \right\|}{t_j^\mathrm{u}[n]} \leq V_{\max}^v, \quad \forall n \in \mathcal{N},
\end{equation}
where $V_{\max}^v$ is the maximum vertical speed. 

\subsubsection{UAV Propulsion Energy Consumption}
The propulsion energy of UAV $j$ at time slot $n$, considering both horizontal and vertical speeds, can be expressed as

\begin{equation}
\begin{aligned}
e_j^\mathrm{u}[n] = & t_j^\mathrm{u}[n]  \left( P_0 \left( 1 + \frac{3(v_j^h[n])^2}{U_{\text{tip}}^2} \right) + \frac{1}{2} d_0 \rho s G \left( v_j^h[n] \right)^3 \right. \\
& \left. + P_1 \left( \sqrt{1 + \frac{(v_j^h[n])^4}{4 v_o^4}} - \frac{(v_j^h[n])^2}{2 v_o^2} \right)^{1/2} + P_2 v_j^v[n] \right)
\end{aligned}
\end{equation}
where $P_0$ and $P_1$ are the constant rotor profile power and induced power during hovering, respectively; $P_2$ is the constant ascent/descent power; $U_{\text{tip}}$ is the tip speed of the rotor blades; $v_0$ is the average rotor-induced speed during hovering; $d_0$ and $s$ are the body drag ratio and rotor solidity, respectively; and $\rho$ and $G$ represent air density and rotor disc area, respectively.

\subsubsection{ RIS Phase Configuration}
As shown in Fig. \ref{fig:model}, a RIS is deployed on the surface of a ground building to redirect the signals between the UAV formation and the GTs, avoiding NLoS connections. Assume the RIS has $M_c \times M_r$ PRUs, forming a uniform planar array (UPA). Specifically, each column of the UPA has $M_c$ PRUs, with a spacing of $d_c$ meters, and each row has $M_r$ PRUs, with a spacing of $d_r$ meters. The position of the first PRU of the RIS in the horizontal dimension is represented as $\mathbf{w}_R = [x_R, y_R]^T$, and in the vertical dimension, it is at $z_R$.

The reflection phase coefficient matrix of the RIS at time slot $n$ is obtained by averaging the phase recommendations of all UAVs: $\overline{\boldsymbol{\Theta}}[n] = \frac {\sum_{j=1}^{J} \boldsymbol{\Theta}_j[n] }{J}$. Specifically, 
\begin{equation}
\boldsymbol{\Theta}_j[n] = \mathrm{diag} (\boldsymbol{\omega} ^{(j)}[n]) \in \mathbb{C}^{M_r M_c \times M_r M_c},
\end{equation}
where 
\begin{equation}
\boldsymbol{\omega}_j[n] = \{ e^{j \omega_{m_r, m_c}^{(j)}[n]} \}_{m_r=1, m_c=1}^{M_r, M_c} \in \mathbb{C}^{M_r M_c \times 1}
\end{equation}
represents the phase parameter vector recommended by UAV $j$ at time slot $n$. $\omega_{m_r, m_c}^{(j)}[n] \in [-\pi, \pi)$ is the phase shift inserted at PRU $(m_r, m_c)$ by UAV $j$ at time slot $n$, for all $m_r \in \{1, 2, \ldots, M_r\}$ and $m_c \in \{1, 2, \ldots, M_c\}$.

\subsubsection{MA on Each UAV}
\begin{figure}[h]
\centering
\includegraphics[width=0.4\textwidth]{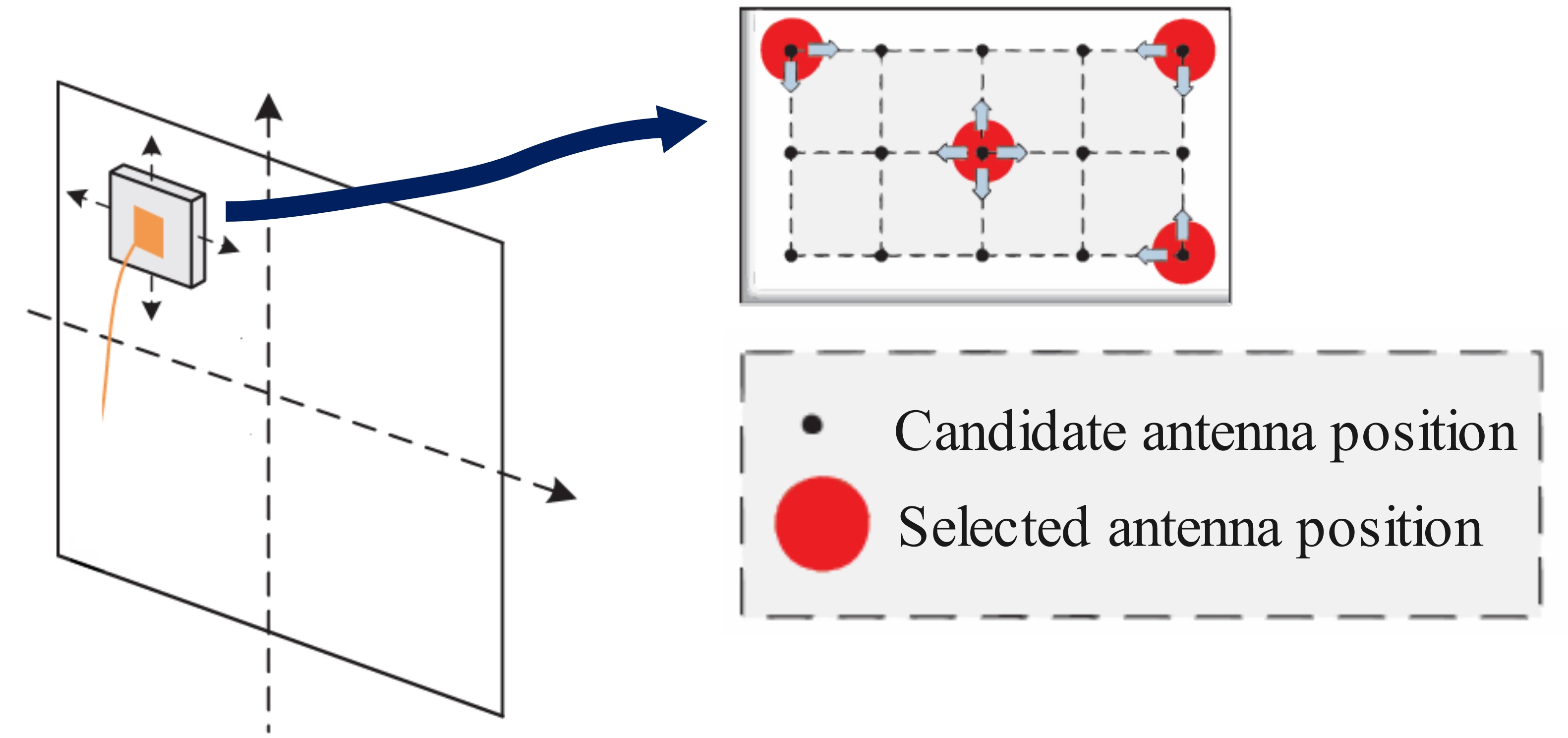} %
\caption{MA positioning with a limited set of discrete locations.}
\label{fig:ma}
\end{figure}
Assume each UAV is equipped with one MA, which is connected to the RF link via a flexible cable controlled by a driving component. The MA can move freely within a limited two-dimensional space $\mathcal{Q}$ consisting of $Q$ discrete positions (see Fig. \ref{fig:ma}).

\begin{equation}
\mathcal{Q} = \left\{ \mathbf{p}_1, \ldots, \mathbf{p}_Q \right\}, \quad \mathbf{p}_q \in \mathbb{R}^{2 \times 1}
\end{equation}

Let $a_{j, q}[n] \in \{0,1\}$ indicate whether the MA of UAV $j$ is at position $\mathbf{p}_q$ at time slot $n$. Here, $\mathbf{p}_q = [\Delta x_q, \Delta y_q]^T$ is the offset relative to the UAV's center. At any given time, the MA of UAV $j$ can only appear at a specific position, i.e.,

\begin{equation}
\sum_{q=1}^{Q} a_{j, q}[n] = 1.
\end{equation}

Thus, the decision variable for the MA positioning is $\mathbf{A}_j[n] = \{a_{j, q}[n]\}_{q=1}^{Q}$.

\subsubsection{ UAV-RIS Link Channel Model}

Assume there are $K$ GTs on the ground, $\mathcal{K} \triangleq \{1, 2, \ldots, K\}$, each in a low-mobility mode. The position of each GT $k$ is represented as $\mathbf{w}_k = [x_k, y_k]^T \in \mathbb{R}^{2 \times 1}$, and $D_k$ is the amount of data that GT $k$ needs to process. The RIS ensures that the UAV-GT link can be replaced by two links: UAV-RIS and RIS-GT, both of which are LoS connections. However, when the UAV-RIS link is severely obstructed by ground obstacles, the signal reflection from the RIS may not guarantee a LoS connection between the UAV and the RIS. This pessimistic scenario is not considered in this paper and will be left for future research.

Since each UAV is equipped with an MA, the actual link geometry should take into account the relative position of the MA. Let the three-dimensional center position of UAV $j$ at time slot $n$ be $\left(x_j^\mathrm{u}[n], y_j^\mathrm{u}[n], H_j^\mathrm{u}[n]\right)$, and the MA's two-dimensional position is selected from the set $\mathcal{Q} = \{\mathbf{p}_1, \ldots, \mathbf{p}_Q\}$. Thus, the actual position of UAV $j$'s MA is:

\begin{equation}
\widetilde{\mathbf{L}}_j^\mathrm{u}[n] = \left[ \begin{array}{c} x_j^\mathrm{u}[n] \\ y_j^\mathrm{u}[n] \end{array} \right] + \sum_{q=1}^{Q} a_{j, q}[n] \cdot \mathbf{p}_q, \quad \widetilde{H}_j^\mathrm{u}[n] = H_j^\mathrm{u}[n].
\end{equation}
Therefore, the true three-dimensional Euclidean distance between UAV $j$ and the RIS is:
\begin{equation}
d_j^\mathrm{ur}[n] = \sqrt{ \left( \widetilde{H}_j^\mathrm{u}[n] - z_R \right)^2 + \left( \widetilde{\mathbf{L}}_j^\mathrm{u}[n] - \mathbf{w}_R \right)^T \left( \widetilde{\mathbf{L}}_j^\mathrm{u}[n] - \mathbf{w}_R \right)}.
\end{equation}

We have the horizontal projection length 
\begin{equation}
r = \sqrt{\left( \widetilde{\mathbf{L}}_j^\mathrm{u}[n] - \mathbf{w}_R \right)^T \left( \widetilde{\mathbf{L}}_j^\mathrm{u}[n] - \mathbf{w}_R \right)},
\end{equation}
Let $\vartheta$ be the horizontal (azimuth) angle measured from the $x$-axis, and let $\theta$ be the elevation angle measured from the horizontal plane.
Define
\begin{equation}
\phi^{\mathrm{ur}}_j[n] \triangleq \cos\vartheta=\frac{\tilde{x}_j^{\mathrm{u}}[n] - x_R}{r}
\end{equation}
\begin{equation}
\varphi^{\mathrm{ur}}_j[n] \triangleq \sin\vartheta= \frac{\tilde{y}_j^{\mathrm{u}}[n] - y_R}{r},
\end{equation}
\begin{equation}
\psi^{\mathrm{ur}}_j[n] \triangleq \cos\theta=\frac{r}{d^{\mathrm{ur}}_j[n]}.
\end{equation}

It follows that
\begin{equation}
\phi^{\mathrm{ur}}_j[n]\psi^{\mathrm{ur}}_j[n]=\cos\vartheta\cos\theta=\frac{\tilde{x}_j^{\mathrm{u}}[n] - x_R}{d^{\mathrm{ur}}_j[n]},
\end{equation}
\begin{equation}
\varphi^{\mathrm{ur}}_j[n]\psi^{\mathrm{ur}}_j[n]=\sin\vartheta\cos\theta=\frac{\tilde{y}_j^{\mathrm{u}}[n] - y_R}{d^{\mathrm{ur}}_j[n]}.
\end{equation}

Hence, $\phi^{\mathrm{ur}}_j[n]\psi^{\mathrm{ur}}_j[n]$ and $\varphi^{\mathrm{ur}}_j[n]\psi^{\mathrm{ur}}_j[n]$ are precisely the direction cosines of the incident unit vector along the RIS row ($x$) and column ($y$) axes.

Let $\lambda$ be the wavelength and $\tfrac{2\pi}{\lambda}$ the wavenumber. 
If the wave advances by $\Delta \ell$ meters, the phase accrues $\tfrac{2\pi}{\lambda}\Delta \ell$.
For a uniform planar RIS aligned with the $xy$-plane, let $d_r$ and $d_c$ denote the inter-element spacings along the row ($x$) and column ($y$) axes, respectively.
Because only the component of the element spacing along the propagation direction contributes to phase, the effective path differences are
\begin{equation}
\Delta \ell_r = d_r/\big(\phi^{\mathrm{ur}}_j[n]\psi^{\mathrm{ur}}_j[n]\big),
\end{equation}
\begin{equation}
\Delta \ell_c = d_c/\big(\varphi^{\mathrm{ur}}_j[n]\psi^{\mathrm{ur}}_j[n]\big),
\end{equation}

\begin{figure}[h]
\centering
\includegraphics[width=0.39\textwidth]{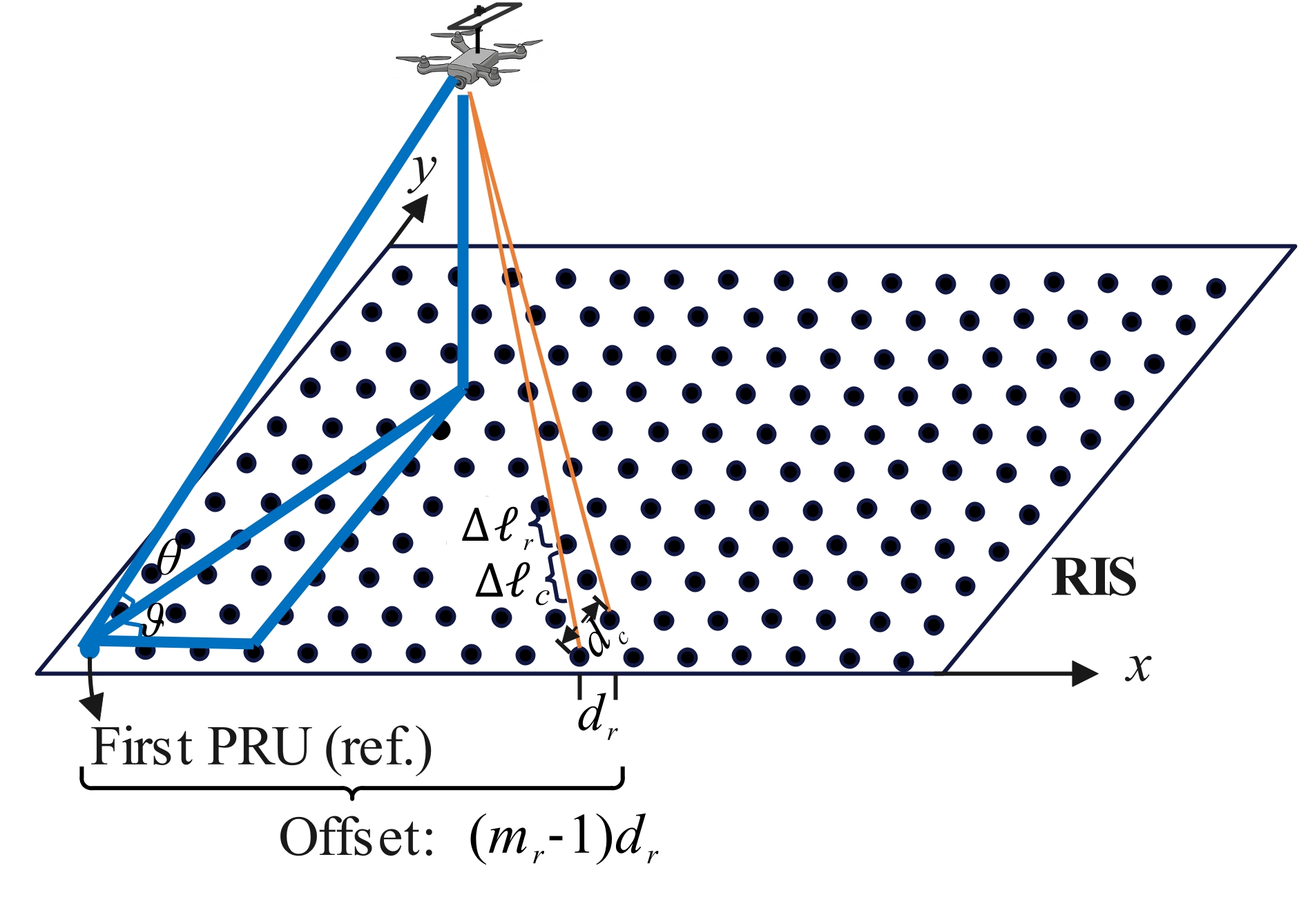} %
\caption{UAV–RIS LoS geometry.}
\label{fig:geo}
\end{figure}
As shown in Fig. \ref{fig:geo},  the first PRU is used as the phase reference. Along the row ($x$) axis, the $m$-th PRU is offset by $(m-1)d_r$ from the reference, where $d_r$ is the inter-element spacing (similarly $d_c$ along the column/$y$ axis). The incident/incoming wave from the UAV impinges with azimuth $\vartheta$ and elevation $\theta$; hence the path difference in the direction of the incoming wave between adjacent PRUs is $\Delta \ell_r =d_r/(\cos\vartheta\cos\theta)= d_r/\big(\phi^{\mathrm{ur}}_j[n]\psi^{\mathrm{ur}}_j[n]\big)$ in the row direction (and $\Delta \ell_c =d_c/(\sin\vartheta\cos\theta)= d_c/\big(\varphi^{\mathrm{ur}}_j[n]\psi^{\mathrm{ur}}_j[n]\big)$ in the column direction), leading to per-element phase steps
\begin{equation}
\Delta \varphi_r =  \frac{2\pi}{\lambda}d_r/\big(\phi^{\mathrm{ur}}_j[n]\psi^{\mathrm{ur}}_j[n]\big),
\end{equation}
\begin{equation}
\Delta \varphi_c =  \frac{2\pi}{\lambda}d_c/\big(\varphi^{\mathrm{ur}}_j[n]\psi^{\mathrm{ur}}_j[n]\big).
\end{equation}

The row- and column-wise steering vectors are
\begin{equation}
\mathbf a_r=\big[1,\,e^{-j\Delta \varphi_r},\,\ldots,\,e^{-j(M_r-1)\Delta \varphi_r}\big]^T,
\end{equation}
\begin{equation}
\mathbf a_c=\big[1,\,e^{-j\Delta \varphi_c},\,\ldots,\,e^{-j(M_c-1)\Delta \varphi_c}\big]^T,
\end{equation}
and the UPA response is $\mathbf a_r\otimes \mathbf a_c$.
Under a narrowband LoS far-field model, the UAV--RIS link takes the form
\begin{equation}
\begin{aligned}
& g_j^\mathrm{ur}[n] = \frac{\sqrt{\xi}}{d_j^\mathrm{ur}[n]} \\
& \left[ 1, e^{-j \frac{2 \pi}{\lambda} d_r/ \left( \phi_j^\mathrm{ur}[n] \cdot \psi_j^\mathrm{ur}[n] \right)}, \ldots, e^{-j \frac{2 \pi}{\lambda} \left( M_r - 1 \right) d_r/ \left( \phi_j^\mathrm{ur}[n] \cdot \psi_j^\mathrm{ur}[n] \right)} \right]^T \otimes \\
&  \left[ 1, e^{-j \frac{2 \pi}{\lambda} d_c/ \left( \varphi_j^\mathrm{ur}[n] \cdot \psi_j^\mathrm{ur}[n] \right)}, \ldots, e^{-j \frac{2 \pi}{\lambda} \left( M_c - 1 \right) d_c /\left( \varphi_j^{ur}[n] \cdot \psi_j^\mathrm{ur}[n] \right)} \right]^T
\end{aligned}
\end{equation}
where $\xi$ is the path-loss constant at the reference distance $D_0=1\,\mathrm{m}$.

\subsubsection{RIS-GT Link Channel Model}

Assume that the RIS is modeled as a far-field array response vector, because $d_j^{\mathrm{ur}}[n] \gg \max \left\{ M_r d_r, M_c d_c \right\}$. Similarly, the channel gain from the RIS to the $k$-th GT is $g_k^\mathrm{rg} \in \mathbb{C}^{M_r \times M_c}$, which is given by
\begin{equation}
\begin{aligned}
g_k^\mathrm{rg} &= \frac{\sqrt{\xi}}{d_k^\mathrm{rg}} \left[ 1, e^{-j \frac{2 \pi}{\lambda} d_r/ (\phi_k^\mathrm{rg} \psi_k^\mathrm{rg})}, \ldots, e^{-j \frac{2 \pi}{\lambda} (M_r - 1) d_r/ (\phi_k^\mathrm{rg} \psi_k^\mathrm{rg})} \right]^T \\
& \otimes \left[ 1, e^{-j \frac{2 \pi}{\lambda} d_c/ (\varphi_k^\mathrm{rg} \psi_k^\mathrm{rg})}, \ldots, e^{-j \frac{2 \pi}{\lambda} (M_c - 1) d_c/ (\varphi_k^\mathrm{rg} \psi_k^\mathrm{rg})} \right]^T 
\end{aligned}
\end{equation}
where $d_k^\mathrm{rg} = \sqrt{z_R^2 + \left( \mathbf{w}_R - \mathbf{w}_k \right)^2}$, $\phi_k^\mathrm{rg} = \frac{x_k - x_R}{||\mathbf{w}_R - \mathbf{w}_k||}$, and $\varphi_k^\mathrm{rg} = \frac{y_k - y_R}{||\mathbf{w}_R - \mathbf{w}_k||}$ represent the cosine and sine of the horizontal angle of departure (AoD) from the RIS to the $k$-th GT in horizontal plane, respectively; $\psi_k^\mathrm{rg} = \frac{||\mathbf{w}_R - \mathbf{w}_k||}{d_k^\mathrm{rg}}$ represents the cosine of the vertical AoD.

\subsubsection{ UAV–RIS–GT Cascaded Channel Model}

\begin{figure*}[h]
\begin{equation}
\begin{aligned}
\min  \sum_{n=1}^{N} \frac{\sum_{j=1}^{J} t_j^u[n]  \left( P_0 \left( 1 + \frac{3 (v_j^h[n])^2}{U_{\mathrm{tip}}^2} \right) + \frac{1}{2} d_0 \rho s G (v_j^h[n])^3 + P_1 \left( \sqrt{1 + \frac{(v_j^h[n])^4}{4 v_o^4}} - \frac{(v_j^h[n])^2}{2 v_o^2} \right)^{1/2} + P_2 v_j^v[n] \right) } {\sum_{k=1}^{K}  t_j^\mathrm{u}[n] r_k[n] } \\
\end{aligned}
\label{eq:prob}
\end{equation}
\noindent\rule{\textwidth}{0.4pt}
\end{figure*}

Based on the above, the cascaded channel gain from UAV $j$ to the $k$-th GT at time slot $n$ can be expressed as

\begin{equation}
g_{j, k}^{\mathrm{urg}}[n]=a\left(g_{k}^\mathrm{rg}\right)^{T}  \overline{\boldsymbol{\Theta}}[n]  g_{j}^\mathrm{ur}
\end{equation}

To evaluate the probability that the direct UAV–GT link is blocked, this paper adopts the air-to-ground channel model for urban environments as in [16]. The blockage probability between UAV $j$ and GT $k$ at time slot $n$ is given by

\begin{equation}
p_{j, k}[n]=1-\frac{1}{1+a \exp \left(-b\left(\arctan \left(\frac{\widetilde{H}_j^\mathrm{u}[n]}{d_{j, k}^\mathrm{ug}}\right)-a\right)\right)}
\end{equation}
where $d_{j, k}^\mathrm{ug} = \sqrt{\left( \widetilde{H}_j^\mathrm{u}[n] \right)^2 + \left\| \widetilde{\mathbf{L}}_j^\mathrm{u}[n] - \mathbf{w}_k \right\|^2 }$, and $a$ and $b$ are environment-dependent constants. Then, the average channel gain and data rate achievable at the $k$-th GT at time slot $n$ can be written as

\begin{equation}
\begin{aligned}
g_k[n]=\underbrace{\sum_{j=1}^{J} (1-p_{j,k}[n]) \frac{\xi}{(d_{j,k}^\mathrm{ug})^2}}_{\text{UAV-GT direct link}} + \underbrace{\sum_{j=1}^{J} p_{j,k}[n] g_{j,k}^\mathrm{urg}[n]}_{\text{UAV-RIS-GT cascaded link}}  
\end{aligned}
\end{equation}
\begin{equation}
\begin{aligned}
r_k[n] = c_k[n] B \log_2 \left( 1 + \frac{g_k[n]P }{B \sigma^2} \right)  
\end{aligned}
\end{equation}
where $P$ denotes the fixed transmit power of the UAV, $B$ is the bandwidth, and $\sigma^2$ is the noise variance. $c_k[n] \in \{0, 1\}$ indicates whether GT $k$ is scheduled at slot $n$. The value of $c_k[n]$ is determined as the mode of all UAVs’ individual scheduling decisions $\{c_{j, k}[n]\}_{j=1}^{J}$:

\begin{equation}
c_k[n] = \operatorname{mode}\left(\left\{c_{j, k}[n]\right\}_{j=1}^{J}\right)
\end{equation}

That is, GT $k$ is selected if the majority of UAVs vote to schedule it. The constraint $\sum_{k=1}^{K} c_k[n] \leq 1, \forall n$ ensures that the RIS serves at most one GT per time slot, following a Time Division Multiple Access (TDMA) protocol. The more complex case where the RIS serves multiple GTs in the same slot via Orthogonal Frequency Division Multiple Access (OFDMA) is left for future work.

\subsection{Problem Formulation}
For each UAV $j$, the following decisions are made at time slot $n$ ($n \in \mathcal{N}$):

\noindent[UAV Trajectory Planning]

Let $a_j^{\text{hor}}[n] \in \{\text{north, south, east, west, hover}\}$ be the horizontal movement decision. The UAV's discrete horizontal position $\mathbf{L}_j^\mathrm{u}[n] = (x^\mathrm{u}_{j}[n], y^\mathrm{u}_{j}[n])$ is updated as:
\begin{equation}
\mathbf{L}_j^\mathrm{u}[n] = 
\begin{cases}
(x^\mathrm{u}_{j}[n{-}1], y^\mathrm{u}_{j}[n{-}1] + l_s), & \text{if } a_j^{\text{hor}}[n] = \text{north}, \\
(x^\mathrm{u}_{j}[n{-}1], y^\mathrm{u}_{j}[n{-}1] - l_s), & \text{if } a_j^{\text{hor}}[n] = \text{south}, \\
(x^\mathrm{u}_{j}[n{-}1] + l_s, y^\mathrm{u}_{j}[n{-}1]), & \text{if } a_j^{\text{hor}}[n] = \text{east}, \\
(x^\mathrm{u}_{j}[n{-}1] - l_s, y^\mathrm{u}_{j}[n{-}1]), & \text{if } a_j^{\text{hor}}[n] = \text{west}, \\
(x^\mathrm{u}_{j}[n{-}1], y^\mathrm{u}_{j}[n{-}1]), & \text{if } a_j^{\text{hor}}[n] = \text{hover}.
\end{cases}
\end{equation}
where $l_s$ denotes the spatial resolution (grid size) in meters.

Let $a_j^{\text{ver}}[n] \in \{\text{ascend, descend, stay}\}$ be the vertical action, which changes the height level as:
\begin{equation}
h_j^\mathrm{u}[n] = 
\begin{cases}
h_j^\mathrm{u}[n{-}1] + 1, & \text{if } a_j^{\text{ver}}[n] = \text{ascend},\\
h_j^\mathrm{u}[n{-}1] - 1, & \text{if } a_j^{\text{ver}}[n] = \text{descend }, \\
h_j^\mathrm{u}[n{-}1], & \text{if } a_j^{\text{ver}}[n] = \text{stay}.
\end{cases}
\end{equation}
Here, the actual UAV altitude is given by $H_j^\mathrm{u}[n] = h_j^\mathrm{u}[n] \cdot h_s$, where $h_s$ is the vertical distance per height level.

\noindent[MA Positioning]
\begin{equation}
\mathbf{A}_j[n] = \{ a_{j, q}[n] \}_{j=1}^J.
\end{equation}

\noindent[GT Scheduling Votes]
\begin{equation}
\mathbf{C}_j[n] = \{ c_{j, k}[n] \}_{j=1}^J.
\end{equation}

\noindent[Slot Duration]
\begin{equation}
t_j^\mathrm{u}[n].
\end{equation}

\noindent[RIS Configuration Recommendation]
\begin{equation}
\boldsymbol{\omega}_j[n] = \{ e^{j \omega_{m_r, m_c}^{(j)}[n]} \}_{m_r=1, m_c=1}^{M_r, M_c}.
\end{equation}

The objective is to minimize the total propulsion energy consumption of all UAVs while maximizing data volume of all GTs  over all time slots, $\mathcal{P}: \min \sum_{n=1}^{N}\frac{\sum_{j=1}^{J}  e_j^\mathrm{u}[n]} {\sum_{k=1}^{K}  t_j^\mathrm{u}[n] r_k[n]} $, which can be detailed as Eq. (\ref{eq:prob}):

Constraints:
\begin{equation}
\begin{aligned}
&\text{C1a}: \quad c_{j, k}[n] \in \{0,1\},\\
&\text{C1b}:  \quad  c_k[n] = \operatorname{mode}(\{c_{j, k}[n]\}_{j=1}^J), \\
&\text{C1c}:  \quad \sum_{k=1}^{K} c_k[n] \leq 1 ; \\
&\text{C2}: \quad \sum_{j=1}^{J} \sum_{n=1}^{N} t_j^\mathrm{u}[n] r_k[n] \geq D_k, \quad \forall k \in \mathcal{K};  \\
&\text{C3}: \quad v_j^h[n] = \frac{ \| L_j^\mathrm{u}[n+1] - L_j^\mathrm{u}[n] \| }{ t_j^\mathrm{u}[n] } \leq V_{\max}^h ; \\
&\text{C4}: \quad v_j^v[n] = \frac{ \| H_j^\mathrm{u}[n+1] - H_j^\mathrm{u}[n] \| }{ t_j^\mathrm{u}[n] } \leq V_{\max}^v ; \\
&\text{C5}: \quad h_{\min} \leq H_j^\mathrm{u}[n] \leq h_{\max} ;\\
&\text{C6}: \quad t_{\min} \leq t_j^\mathrm{u}[n] \leq t_{\max} ;\\
&\text{C7}: \quad \omega_{m_r, m_c}^{(j)}[n] \in [-\pi, \pi) .
\end{aligned}
\end{equation}
where
\begin{itemize}
\item  (C1) ensures that the RIS serves at most one GT in each time slot (TDMA mode);

\item  (C2) ensures that each GT’s data demand $D_k$ is completed within the total mission time of the UAV fleet;

\item  (C3), (C4), and (C5) are kinematic constraints for rotary-wing UAVs;

\item  (C6) constrains the working duration of UAV $j$ in slot $n$ within specified bounds;

\item  (C7) restricts the recommended RIS phase to the interval $[-\pi, \pi)$.
\end{itemize}

\section{Solved by MARL}

\subsection{Networked MDP}
\begin{definition}[Networked MDP]
We model the $J$ cooperative UAVs as a networked cooperative multi-agent MDP: 
\begin{equation}
\left\{G, \left\{ \mathcal{S}_j, \mathcal{A}_j \right\}_{j \in \mathcal{V}}, \left\{ \mathcal{M}_{ij} \right\}_{ij \in \mathcal{E}}, p, \left\{ \mathcal{R}_j \right\}_{j \in \mathcal{V}} \right\},
\end{equation}
with communication graph $G(\mathcal{V}, \mathcal{E})$, the augmented state space of each UAV $j$ is given by its local UAV state $\mathcal{S}_j[n]$ and the states of neighboring UAVs:
$
\tilde{\mathcal{S}}_{j}[n] := \mathcal{S}_{j}[n] \cup \left\{ \mathcal{S}_{i}[n] \right\}_{i \in \mathcal{N}_j}
$ via neighbor messages $\{ \mathcal{M}_{ij} \}_{i \in \mathcal{N}_j}$, where $\mathcal{N}_j := \{ i \in \mathcal{V} |\| \mathbf{L}_i^\mathrm{u}[n-1] - \mathbf{L}_j^\mathrm{u}[n-1]\|^2 + | h_i^\mathrm{u}[n-1] - h_j^\mathrm{u}[n-1] |^2 \leq h^2\}$ denotes the set of UAVs within radius (hop distance) $h$ of UAV $j$. The global reward is defined as the average of all local rewards:
$
\mathcal{R}[n] = \frac{1}{|\mathcal{V}|} \sum_{j \in \mathcal{V}} \mathcal{R}_j[n], \mathcal{V} = \{1, \ldots, J\}
$. The state transition depends on the states and actions of $j$ and its neighbors:
\begin{equation}
p_j\big( \mathcal{S}_j[n+1] \mid \mathcal{S}_{\mathcal{V}_j}[n], \mathcal{A}_j[n], \mathcal{A}_{\mathcal{N}_j}[n] \big), p_j \in p
\end{equation}
where $\mathcal{V}_j=\mathcal{N}_j\cup\{j\}$.  
\end{definition}

\subsubsection{State Space}
The local state of UAV $j$ in time slot $n$ is defined as:
\begin{equation}
\mathcal{S}_j[n] = \left( \mathbf{L}_j^\mathrm{u}[n-1], h_j^\mathrm{u}[n-1] \right).
\label{eq:state}
\end{equation}
where $\mathbf{L}_j^\mathrm{u}[n-1]$ denotes the horizontal location on a discretized grid at time slot $n-1$, and $h_j^\mathrm{u}[n-1]$ is the altitude level (discretized) at time slot $n-1$. 

\subsubsection{Action Space}
The action comprises trajectory update, MA positioning, GT scheduling, slot duration, and RIS phase configuration:
\begin{equation}
\mathcal{A}_j[n] = \left( a_j^{\text{hor}}[n], a_j^{\text{ver}}[n], \mathbf{A}_j[n], \mathbf{C}_j[n], t_j^\mathrm{u}[n], \boldsymbol{\omega}_j[n] \right).
\label{eq:action}
\end{equation}
Use four categorical (\texttt{softmax}) heads for the discrete variables
$a_j^{\mathrm{hor}}[n],\, a_j^{\mathrm{ver}}[n],\, \mathbf{A}_j[n],\, \mathbf{C}_j[n]$; sample each to an one-hot vector and concatenate them.
Output the scalar continuous variable $t_j^{\mathrm{u}}[n]$.
Output the multi-dimensional continuous variable $\boldsymbol{\omega}_j[n]=\{e^{i\omega_{m_r,m_c}^{(j)}[n]}\}$.
Finally, concatenate all six components to form final action vector.

\subsubsection{Reward}
The reward at each time step is defined as:
\begin{equation}
\mathcal{R}_j[n] = \frac{\sum_{k=1}^{K}  t_j^\mathrm{u}[n] r_k[n]} { e_j^\mathrm{u}[n]}
\end{equation}
which is basically the ratio of the total amount of data of GTs up to time slot $n$ to the UAV's propulsion energy.

\subsection{Spatiotemporal RL}
\begin{definition}[Spatiotemporal Discounted Reward]
The spatiotemporally discounted reward of UAV $j$ in time slot $n$ is 
\begin{equation}
\hat{\mathcal{R}}_j[n] = \sum_{m=n}^{n_\mathcal{B}-1} \gamma^{m-n} \sum_{i\in\mathcal{V}} \alpha^{d_{ji}} \mathcal{R}_i[m],
\end{equation}
where $d_{ji}$ is the shortest path length between UAVs $j$ and $i$, $\alpha\in[0,1]$ is the spatial discount factor, $\gamma$ the temporal discount factor, and $\mathcal{B}$ denotes the set of indices in the sampled minibatch (of size $|\mathcal{B}|$), with $n_\mathcal{B}$ being the last index.
\end{definition}

We adopt the Advantage Actor-Critic (A2C) method.  The actor network loss for UAV $j$ is:
\begin{equation}
\begin{aligned}
\mathcal{L}(\theta_j) = &\frac{1}{|\mathcal{B}|} \sum [ -\log \pi_{\theta_j}(\mathcal{A}_j[n] \mid \tilde{\mathcal{S}}_j[n]) \Delta_j[n] \\
&+ \beta  \mathrm{H}\big(\pi_{\theta_j}(\mathcal{A}_j[n]\mid\tilde{\mathcal{S}}_j[n])\big) ],
\end{aligned}
\end{equation}
where $\Delta_j[n] = \hat{\mathcal{R}}_j[n] - V_{\phi_j}(\tilde{\mathcal{S}}_j[n],\mathcal{A}_{\mathcal{N}_j}[n])$ is the advantage function of UAV $j$, $V_{\phi_j}(\cdot)$ is the value function (critic) of UAV $j$ with parameters $\phi_j$, $\beta$ is the entropy regularization weight, and $\mathrm{H}\big(\pi_{\theta_j}(\mathcal{A}_j[n]\mid\tilde{\mathcal{S}}_j[n])\big)=-\sum \pi_{\theta_j}(\mathcal{A}_j[n] \mid \tilde{\mathcal{S}}_j[n]) \log \pi_{\theta_j}(\mathcal{A}_j[n] \mid \tilde{\mathcal{S}}_j[n])$ is the policy entropy of all possible actions. The critic network loss is:
\begin{equation}
\begin{aligned}
\mathcal{L}(\phi_j) &=\frac{1}{|\mathcal{B}|} \sum \big[ \Delta_j[n] \big]^2 \\
                           &= \frac{1}{|\mathcal{B}|} \sum \big[ \hat{\mathcal{R}}_j[n] - V_{\phi_j}(\tilde{\mathcal{S}}_j[n],\mathcal{A}_{\mathcal{N}_j}[n]) \big]^2.
\end{aligned}
\end{equation}

\begin{proposition}
Each augmented observation $\tilde{\mathcal{S}}_j[n]$ contains UAV $j$’s local state and aggregated multi-hop information from its own neighbors, recursively propagated up to $h$ hops, while the spatial discount $\alpha^{d_{ji}}$ reduces the influence of distant UAVs, ensuring the learning target is dominated by local interactions spatiotemporally.
\end{proposition}
The proof is provided in Appendix A.

\subsection{Spatiotemporal RL with  Differentiable Communication}
\label{sec:neurcomm}
At the beginning of slot \(n\), all UAVs proceed synchronously as follows:
\begin{enumerate}[label=Step\arabic*:]
\item UAV \(j\) observes the current local state \(\mathcal{S}_j[n]\) and uses the previous-slot policy fingerprint \(\pi_{\theta_j}[n{-}1]\) and belief \(h_j[n{-}1]\) (prior-decision quantities).
\item UAV \(j\) broadcasts an identical message to all neighbors:
\begin{equation}
\mathcal{M}_j[n] \;=\; \mathcal{S}_j[n] \;\cup\; \pi_{\theta_j}[n{-}1] \;\cup\; h_j[n{-}1],
\end{equation}
and receives \(\{\mathcal{M}_i[n]\}_{i\in\mathcal{N}_j}\) from its one-hop neighbors within the same slot.
\item With differentiable encoders \(e^{s}_{\lambda_j}(\cdot), e^{p}_{\lambda_j}(\cdot), e^{h}_{\lambda_j}(\cdot)\) for state, policy, and belief, and a differentiable extractor \(g_{\nu_j}(\cdot)\), UAV \(j\) updates its prior-decision belief:
\begin{equation}
\label{eq:neurcomm-belief}
\begin{aligned}
h_j[n]= g_{\nu_j}(
h_j[n{-}1],\;
&e^{s}_{\lambda_j}\big(\{\mathcal{S}_i[n]\}_{i\in\mathcal{V}_j}\big),\\
&e^{p}_{\lambda_j}\big(\{\pi_{\theta_i}[n{-}1]\}_{i\in\mathcal{N}_j}\big),\\
&e^{h}_{\lambda_j}\big(\{h_i[n{-}1]\}_{i\in\mathcal{N}_j})
),
\end{aligned}
\end{equation}
where \(\mathcal{V}_j=\mathcal{N}_j\cup\{j\}\). Optional attentions can be applied at the sender or receiver,
\(\mu^{\mathrm{send}}_j(\mathcal{M}_j[n])\) or \(\mu^{\mathrm{recv}}_j(\{\mathcal{M}_i[n]\}_{i\in\mathcal{N}_j})\), both folded into \eqref{eq:neurcomm-belief}.
\end{enumerate}

\begin{proposition}[Spatial information propagation]
\label{prop:neurcomm}
Under the prior-decision timing with one-hop-per-slot latency, the earliest information from UAV \(i\) that can affect \(h_j[n]\) obeys
\begin{equation}
\label{eq:delayed-global}
h_j[n] \supset \mathcal{S}_j[0{:}n] \;\cup\; \Big\{\mathcal{S}_i[0{:}n{+}1{-}d_{ji}],\; \pi_{\theta_i}[0{:}n{-}d_{ji}]\Big\}_{i\in\mathcal{V}\setminus j},
\end{equation}
where $x \supset y$ means ``$y$ is utilized to estimate $x$'', i.e., state information from \(d_{ji}\)-hop neighbors arrives with a delay of \(d_{ji}{-}1\) slots, while policy fingerprints are already delayed by one slot in the message and thus appear with \(d_{ji}\) slots total. 
\end{proposition}
The proof is provided in Appendix B.

Replacing \(\tilde{\mathcal{S}}_j[n]\) by \(h_j[n]\) in A2C gives
\begin{equation}
\begin{aligned}
\mathcal{L}(\theta_j)
&= \frac{1}{|\mathcal{B}|}\sum_{n\in\mathcal{B}}
[-\log \pi_{\theta_j}(\mathcal{A}_j[n]\mid h_j[n])\,\hat{A}_j[n]\\
&+\beta\,\mathrm{H}\big(\pi_{\theta_j}(\cdot\mid h_j[n])\big)],
\end{aligned}
\end{equation}

\begin{equation}
\begin{aligned}
\mathcal{L}(\phi_j)
= \frac{1}{|\mathcal{B}|}\sum_{n\in\mathcal{B}}
\Big[\hat{\mathcal{R}}_j[n]-V_{\phi_j}(h_j[n],\mathcal{A}_{\mathcal{N}_j}[n])\Big]^2,
\end{aligned}
\end{equation}
where \(\hat{A}_j[n]=\hat{\mathcal{R}}_j[n]-V_{\phi_j}(h_j[n],\mathcal{A}_{\mathcal{N}_j}[n])\).
Execution remains fully decentralized since \(g_{\nu_j}\) depends only on locally received neighbor messages.

\begin{proposition}[Spatial gradient propagation]
\label{prop:grad}
Under the prior-decision message propagation with one-hop-per-slot latency and differentiable communication/update maps, for any distinct UAVs \(i\neq j\) and any slot \(\tau\),
\begin{equation}
\mathcal{M}_i[\tau]\in\mathrm{anc}\!\big(h_j[\tau+d_{ji}-1]\big),
\end{equation}
i.e., the first time \(\mathcal{M}_i[\tau]\) can influence \(h_j[\cdot]\) is \(\tau+d_{ji}-1\). 
\end{proposition}
The proof is provided in Appendix C.

Because of network delay, a UAV only get a training signal from another UAV’s loss after the message has passed through the network. The first few samples—equal to the number of hops between them minus one—contribute nothing. 

\begin{figure}[h]
\centering
\includegraphics[width=0.5\textwidth]{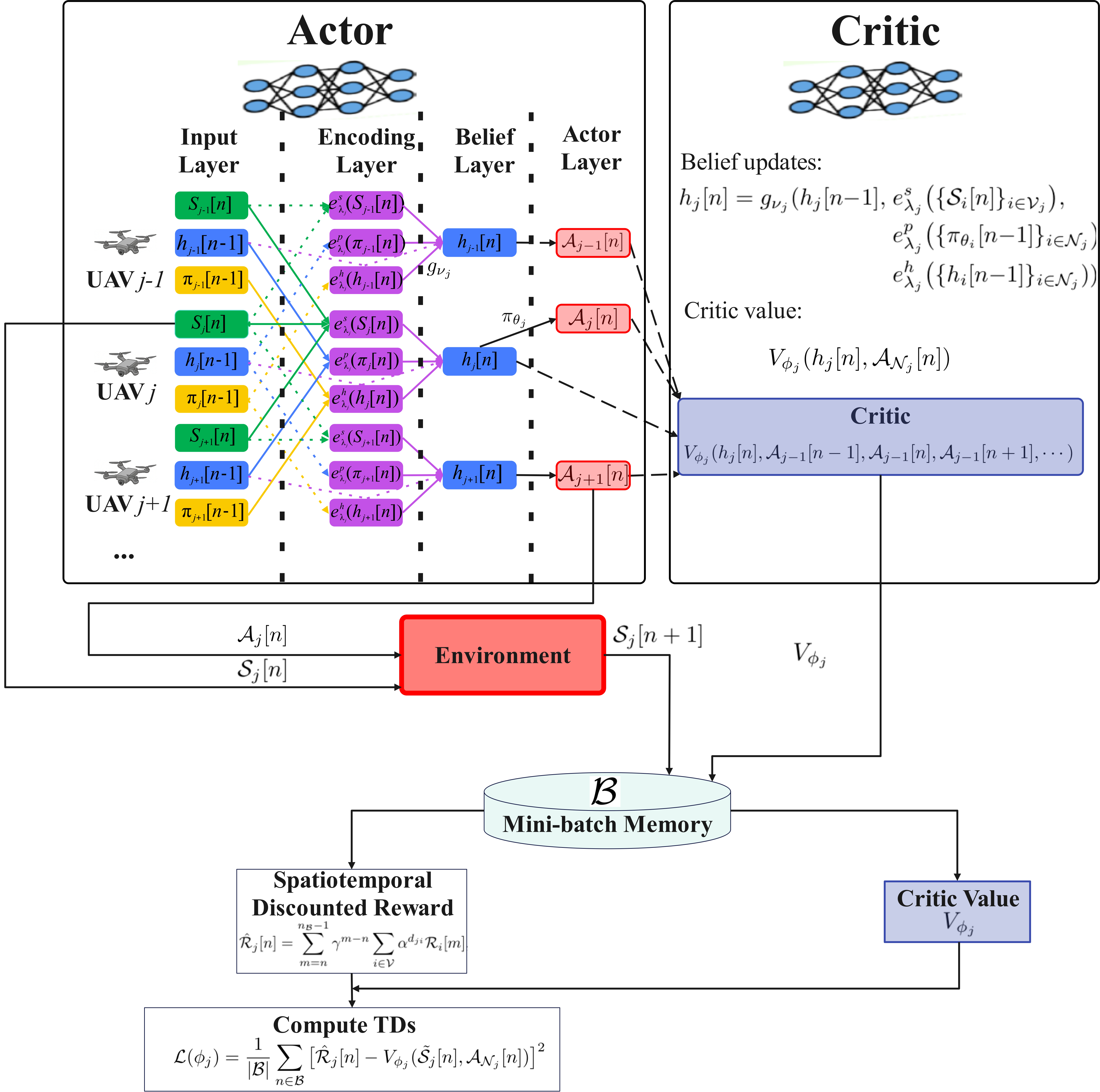}
\caption{Framework of our method.}\label{a2c}
\end{figure}

Our framework (Fig. \ref{a2c}) lets UAVs update beliefs from local and neighbor data, choose actions with the actor, and estimate values with the critic. Rewards are used to compute spatiotemporal returns and the temporal-difference (TD) loss.

\section{Experiments}\label{sec:exp}

\subsection{Experimental Setup}
Environment settings are summarized in Table \ref{tab:env}, and the MARL model and training hyperparameters are listed in Table \ref{tab:marl}. All experiments were conducted on an HPC cluster with two 24-core Intel Xeon Scalable Cascade Lake 8168 CPUs (2.7 GHz), 1.5 TB of DDR4-2666 ECC memory, and sixteen NVIDIA Tesla V100 GPUs.
\begin{figure}[h]
\centering
\includegraphics[width=0.24\textwidth]{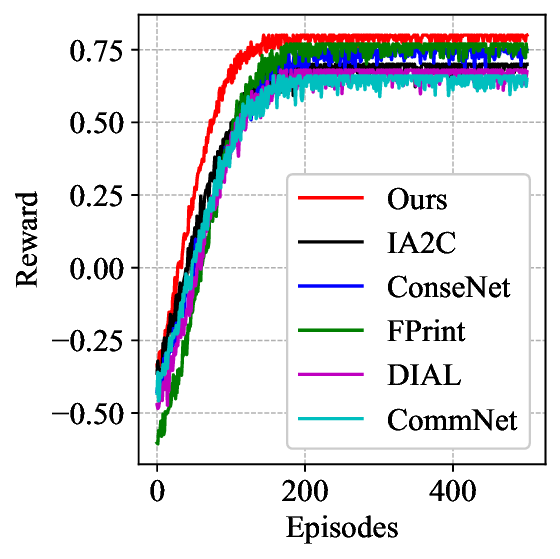}
\caption{Reward (avg. of all UAVs).}\label{fig:reward}
\end{figure}

\begin{figure}[htbp]
    \centering
    \begin{subfigure}[t]{0.24\textwidth}
        \includegraphics[width=\textwidth]{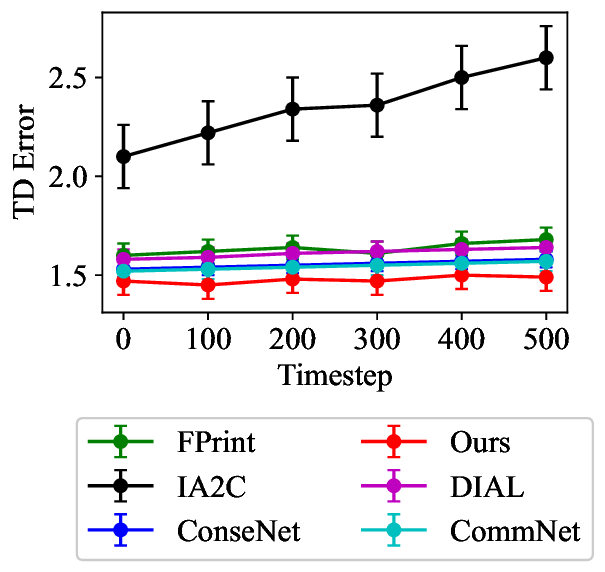}
        \caption{TD error ($\alpha=0.8$).}
    \end{subfigure}
    \hfill
    \begin{subfigure}[t]{0.24\textwidth}
        \includegraphics[width=\textwidth]{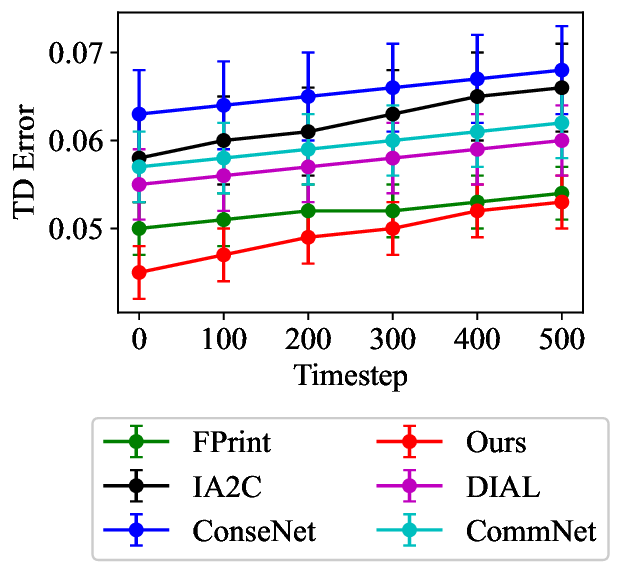}
        \caption{TD error ($\alpha=0.9$).}
    \end{subfigure}
    \hfill
    \begin{subfigure}[t]{0.24\textwidth}
        \includegraphics[width=\textwidth]{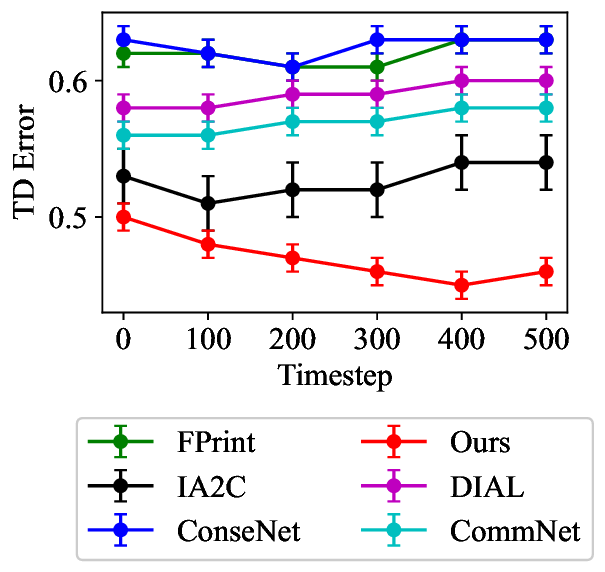}
        \caption{TD error ($\alpha=1.0$).}
    \end{subfigure}
    \hfill
    \begin{subfigure}[t]{0.24\textwidth}
        \includegraphics[width=\textwidth]{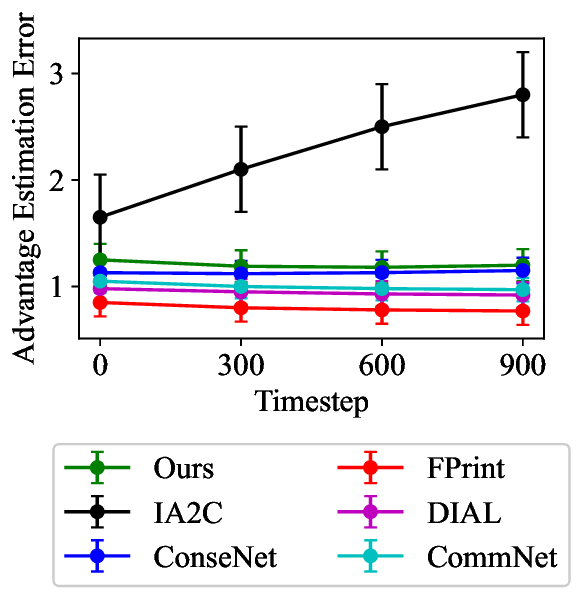}
        \caption{Advantage estimation error ($\alpha=0.8$).}
    \end{subfigure}
    \hfill
    \begin{subfigure}[t]{0.24\textwidth}
        \includegraphics[width=\textwidth]{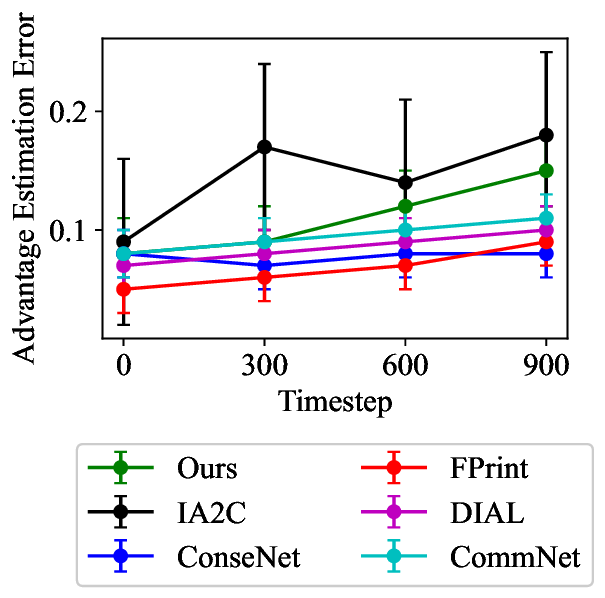}
        \caption{Advantage estimation error ($\alpha=0.9$).}
    \end{subfigure}
    \hfill
    \begin{subfigure}[t]{0.24\textwidth}
        \includegraphics[width=\textwidth]{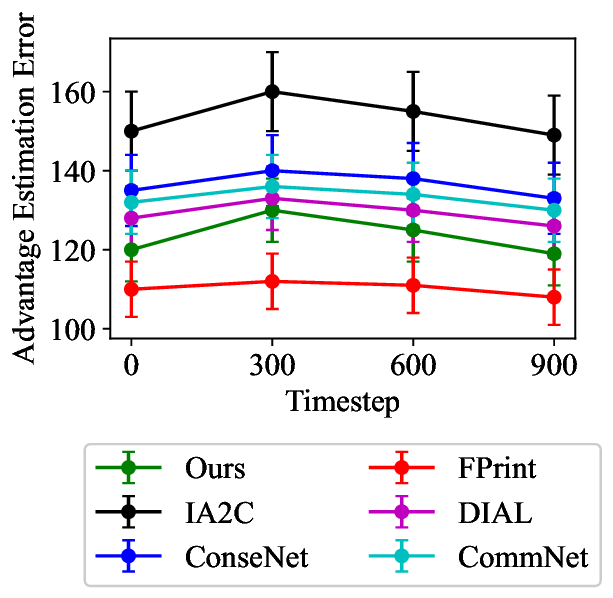}
        \caption{Advantage estimation error ($\alpha=1.0$).}
    \end{subfigure}
    \caption{Critic error  (avg. of all UAVs).}
    \label{fig:error}
\end{figure}

\begin{table}[t]
\centering
\caption{Environment parameters}
\label{tab:env}
\begin{tabular}{l l}
\hline
Parameter & Value\\
\hline
 $J$, $K$ & $10$, $6$ \\
AoI  & $100{\times}100$\,cells \\
$l_s$ & $10$\,m/cell\\
$h$ & $10$\,m (one hop) \\
$\mathbf{L}_0^\mathrm{u},\mathbf{L}_f^\mathrm{u}$,$h_0^{\mathrm{u}}$ & $[0,0]^T$, $[100,100]^T$ cells, 30 levels\\
$\mathbf{w}_R$, $z_R$ & $[50,50]^T$ cells, $50$ levels \\
$h_s$ & $2$\,m/level \\
$[h_{\min},h_{\max}]$ & $[30, 100]$ levels \\
$V_{\max}^h,V_{\max}^v$ & 10m/s, 10m/s \\
$t_{\min}$, $t_{\max}$ & $1$s, $3$s; \\
$N$ & $60$  \\
$B$, $P$ & $2$\,MHz,\; $500$\,mW \\
$\sigma$ & $-169$\,dBm/Hz \\
$a,b,\xi$ & $9.61,\,0.16, \, 10^{0.3}$\\
$U_{\text{tip}},v_0,d_0,s,\rho,G$ & 120,4.3,0.6,0.05,1.225,0.503 \\
$P_0,P_1,P_2$ & $\frac{12 \times 30^3\times 0.4^3\times1.225\times0.05\times0.503}{8},\ \frac{1.1 \times 20^{3/2}}{\sqrt{2 \times 1.225 \times 0.503}}, 11.46$ \\
$D_k$ & 512 Kb \\
$\lambda$ & 0.1 m\\
$M_c\times M_r$; $d_c,d_r$ & $16{\times}16$; $\lambda/2$,$\lambda/2$ \\
$M;\,\theta_i$ & $20;\ 0^\circ$ \\
$Q$ &  $9$ (3$\times$3) \\
$\Delta x_q,\Delta y_q$& $\frac{\lambda}{2}$\\
\hline
\end{tabular}
\end{table}

\begin{table}[t]
\centering
\caption{MARL parameters}
\label{tab:marl}
\begin{tabular}{l l}
\hline
Parameter & Value\\
\hline
$e^{s}_{\lambda_j},\, e^{p}_{\lambda_j},\, e^{h}_{\lambda_j}$ & \texttt{ReLU} \\
$g_{\nu_j}$ & \texttt{LSTM} \\
$\gamma$ & $0.99$ \\
$\beta$ & $0.005$ \\
$\alpha$ & $\{0.8,\,0.9,\,1.0\}$ \\
$\eta_{\text{actor}}$ & $5{\times}10^{-4}$ (\texttt{Adam}) \\
$\eta_{\text{critic}}$ & $2.5{\times}10^{-4}$ (\texttt{Adam}) \\
$|\mathcal{B}|$ & $120$ steps (rollout length) \\
\hline
\end{tabular}
\end{table}

{Baselines}:
\begin{itemize}
  \item \textbf{IA2C}\cite{lowe2017multi}: $h_{i}[n]=\operatorname{LSTM}(h_{i}[n-1], \operatorname{relu}\left(\mathcal{S}_{i}[n]\right))$.
  \item \textbf{ConseNet}\cite{zhang2018fully}: same as IA2C but with consensus critic update. The actor and critic are $\pi_{i}[n]=\operatorname{softmax}\left(h_{i}[n]\right)$, and $V_{i}[n]=\operatorname{linear}\left(\operatorname{concat}\left(h_{i}[n], \operatorname{onehot}\left(\mathcal{A}_{\mathcal{N}_{i}}[n]\right)\right)\right)$.
  \item \textbf{FPrint}\cite{foerster2017stabilising}: $h_{i}[n]=\operatorname{LSTM}\left(h_{i}[n]\right.$, concat $\left(\operatorname{relu}\left(\mathcal{S}_{i}[n]\right)\right.$, $\operatorname{relu}\left.\left.\left(\pi_{\mathcal{N}_{i}}[n-1]\right)\right)\right)$.
  \item \textbf{CommNet}\cite{sukhbaatar2016learning}: $h_{i}[n]=\operatorname{LSTM}(h_{i}[n-1]$, $\tanh (\mathcal{S}_{\mathcal{V}_{i}}[n])$ $+\operatorname{linear}(\operatorname{mean}(h_{\mathcal{N}_{i}}[n-1])))$.
  \item \textbf{DIAL}\cite{foerster2016learning}: $h_{i}[n]=\operatorname{LSTM}(h_{i}[n-1], \operatorname{relu}(\mathcal{S}_{\mathcal{V}_{i}}[n])+\operatorname{relu}(\operatorname{relu}(h_{i}[n-1]))+\operatorname{onehot}(\mathcal{A}_{i}[n-1]))$.
  \item \textbf{Ours}: $h_{i}[n]=\operatorname{LSTM}(h_{i}[n-1], \operatorname{concat}(\operatorname{relu}(\mathcal{S}_{\mathcal{V}_{i}}[n])$, $\operatorname{relu}(\pi_{\mathcal{N}_{i}}[n-1]), \operatorname{relu}(h_{\mathcal{N}_{i}}[n-1])))$.
\end{itemize}

\begin{figure}[htbp]
    \centering
    \begin{subfigure}[t]{0.24\textwidth}
        \includegraphics[width=\textwidth]{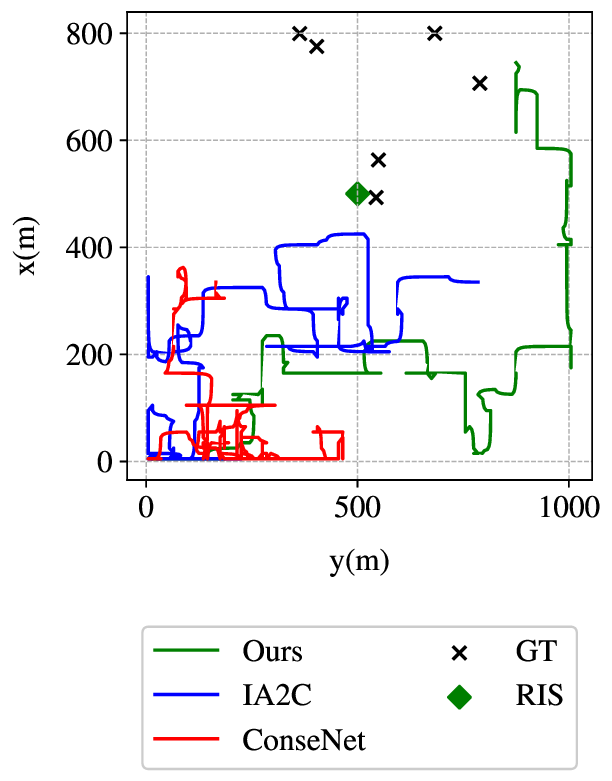}
        \caption{2D trajectory, UAV1.}
        \label{througput}
    \end{subfigure}
    \hfill
    \begin{subfigure}[t]{0.24\textwidth}
        \includegraphics[width=\textwidth]{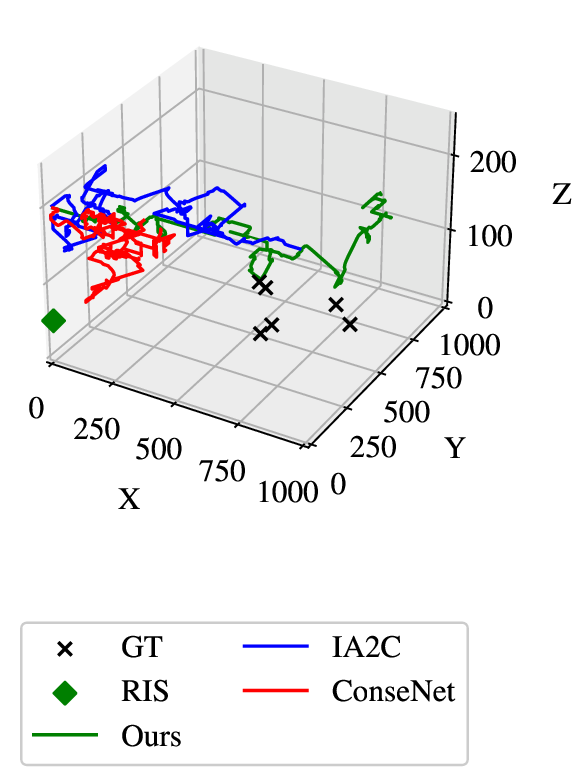}
        \caption{3D trajectory, UAV1.}
        \label{energy}
    \end{subfigure}
   \caption{Trajectory of UAV1.}
    \label{fig:trajectory}
\end{figure}
\begin{figure*}[htbp]
    \centering
    \begin{subfigure}[t]{0.24\textwidth}
        \includegraphics[width=\textwidth]{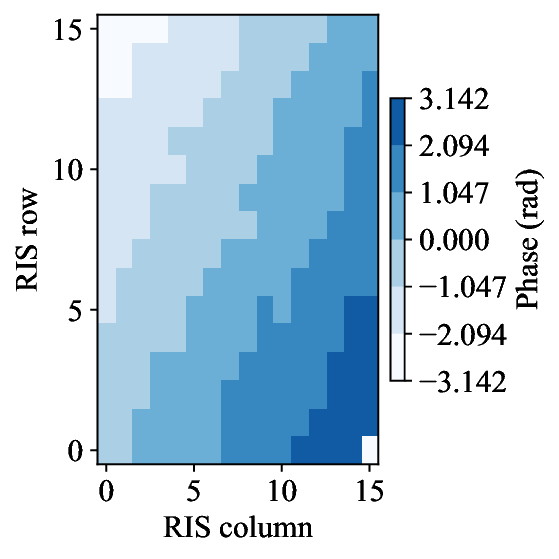}
        \caption{RIS phase in slot 30.}
        \label{fig:ris}
    \end{subfigure}
    \hfill
    \begin{subfigure}[t]{0.24\textwidth}
        \includegraphics[width=\textwidth]{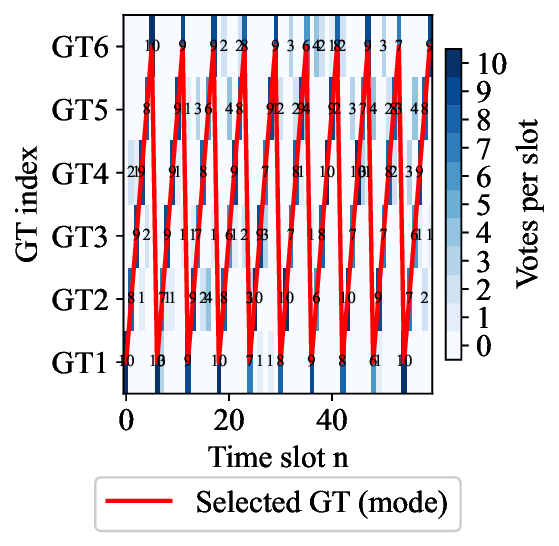}
        \caption{GT voting per slot.}
        \label{fig:gt}
    \end{subfigure}
        \begin{subfigure}[t]{0.24\textwidth}
        \includegraphics[width=\textwidth]{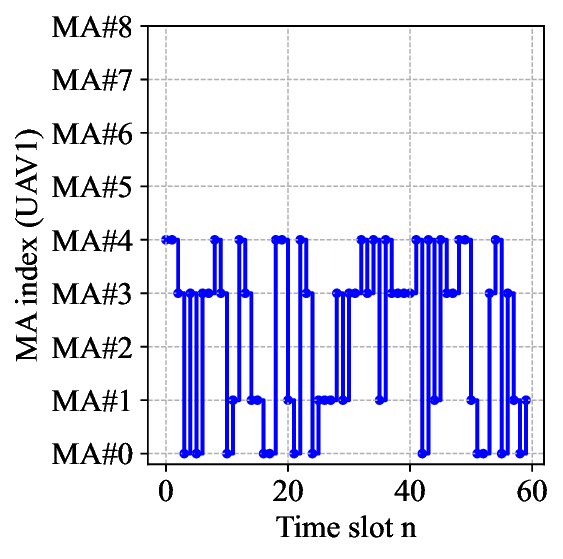}
        \caption{UAV1 MA selection over time.}
        \label{fig:ma1}
    \end{subfigure}
    \hfill
    \begin{subfigure}[t]{0.24\textwidth}
        \includegraphics[width=\textwidth]{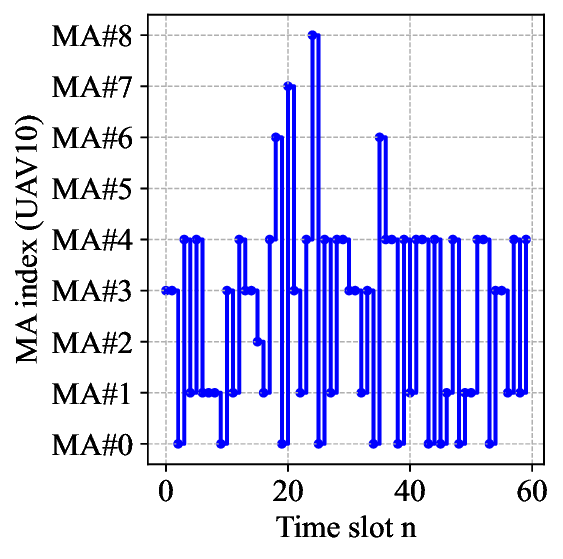}
        \caption{UAV10 MA selection over time.}
        \label{fig:ma2}
    \end{subfigure}
    \caption{RIS recommendation, GTs voting. and MA positioning decisions.}
    \label{fig:decision}
\end{figure*}
\begin{figure}[htbp]
    \centering
    \begin{subfigure}[t]{0.24\textwidth}
        \includegraphics[width=\textwidth]{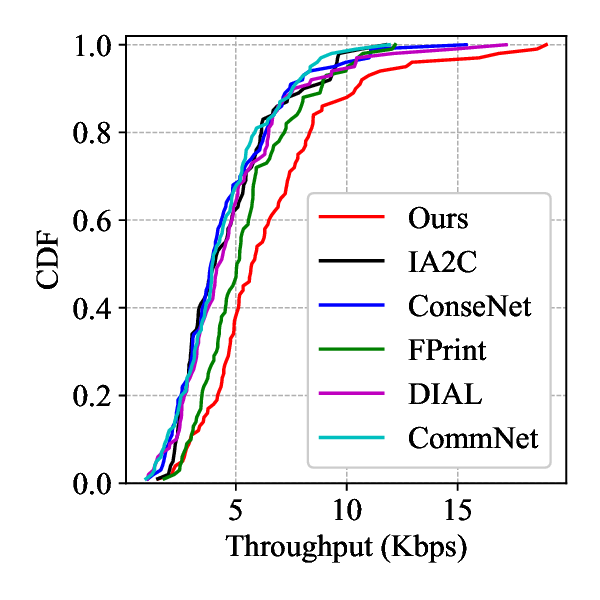}
        \caption{Throughput (Kbps).}
        \label{througput}
    \end{subfigure}
    \hfill
    \begin{subfigure}[t]{0.24\textwidth}
        \includegraphics[width=\textwidth]{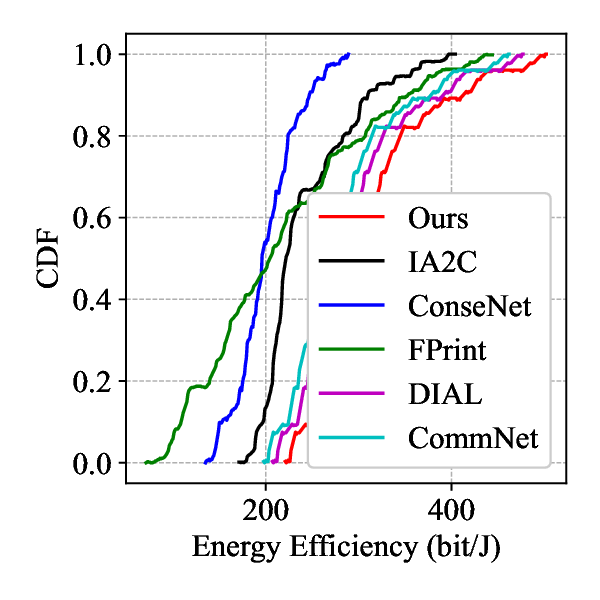}
        \caption{Propulsion Energy (KJ).}
        \label{energy}
    \end{subfigure}
    \caption{Performance comparison for all UAVs (total).}
    \label{fig:perf}
\end{figure}

\subsection{Experimental Results}
\subsubsection{Model Performance}
Across training episodes (Fig. \ref{fig:reward}), our spatiotemporal A2C with differentiable communication (``Ours'') learns much faster and converges to the highest long-run reward among all baselines. The initial rise is steep: by sharing state, policy fingerprints, and beliefs, neighboring UAVs coordinate trajectories, scheduling, slot timing, and RIS phases earlier, yielding more bits per unit of propulsion energy. Methods that either lack explicit multi-hop belief propagation (CommNet/DIAL/FPrint) or transmit only partial summaries (IA2C/ConseNet) plateau lower and learn more noisily, indicating residual non-stationarity.

The critic diagnostics (Fig. \ref{fig:error}) clarify this gap. For all spatial discounts $\alpha\in\{0.8,0.9,1.0\}$, our approach achieves the lowest TD error and advantage-estimation error, with tight error bars—signaling a more stable and accurate value function. As $\alpha$ approaches 1.0 (placing more weight on distant UAVs), errors rise for every method, yet ours remains clearly best. This supports two points: (i) spatial discounting limits long-range influence and reduces variance, and (ii) prior-decision message passing mitigates non-stationarity by aligning updates with one-hop-per-slot propagation. Better critic targets yield better policies, as reflected in the superior reward curve.

\subsubsection{System Performance}
By explicitly augmenting each agent’s local state with the exact 3D coordinates of its neighbors, our policy counts on $\tilde{\mathcal S}_j[n]$ rather than $\mathcal S_j[n])$ as in IA2C/ConseNet, which ignore neighbors. This single change alleviates partial observability at the agent level—since the transition depends on neighbors, providing their positions makes the local state closer to sufficient—so the policy no longer needs exploratory zig-zags to hedge against unknown neighbor motions and can pursue near-shortest routes between tasks, directly lowering the propulsion-energy objective in Eq. \eqref{eq:prob}. 
Knowing who is close to which GT also enables conflict avoidance and division of labor under TDMA (C1): each UAV’s GT-scheduling vote targets unclaimed GTs, preventing redundant visits and reducing travel time needed to satisfy data-demand constraints (C2). In 3D motion planning, awareness of neighbors’ altitudes  informs whether to dive/ascend, yielding fewer aggressive climbs and smoother horizontal decisions $(a_j^{\text{ver}},a_j^{\text{hor}})$ within kinematic limits (C3–C5); at the same time, slot durations $t_j^u[n]$ are chosen more judiciously (C6)—short when a neighbor will take over, longer when $j$ is the only viable server—improving the reward. 
Finally, neighbor geometry serves as side information for RIS/MA coordination: with relative bearings to other UAVs and their served GTs, RIS phase recommendation and MA positioning can be steered to the intended GT while avoiding beam overlap (still respecting C1 and phase bounds C7), which boosts throughput $r_k[n]$ without extra motion. 
Empirically this appears exactly as in the trajectory plots (Fig. \ref{fig:trajectory}): our method, armed with neighborhood states, tends to fly close to the straight-line path and only dives locally when it is the best-positioned server for a GT, whereas IA2C/ConseNet—lacking neighbor coordinates—exhibit myopic detours and unnecessary altitude oscillations, leading to longer paths and higher propulsion energy.

Fig. \ref{fig:decision} summarizes other instantaneous UAV actions.
\begin{enumerate}[leftmargin=*,label=(\alph*)]
\item 
Subfigure \ref{fig:ris} exhibits a clear linear phase gradient across the RIS—monotonically increasing along both rows and columns—which indicates that all PRUs steer the reflected wave toward a single spatial direction. Because the plotted phase is the average recommendation aggregated from all 10 UAVs, the coherent slope reveals strong inter-UAV consensus on the desired reflection direction; small apparent discontinuities are simply due to modulo-$2\pi$ wrapping. This directional beamforming pattern is consistent with a far-field plane-wave model, in which the per-row and per-column phase steps  are proportional to the direction cosines of the target direction, mapping the slope to a unique look angle. The resulting array factor implies a narrow main lobe (with sidelobes commensurate with a $16\times16$ aperture), confirming that the fleet recommends a common steering vector to maximize the cascaded link gain toward the scheduled GT. 
\item In Subfigure~\ref{fig:gt}, votes within each time slot are strongly concentrated on a single GT (the darkest bar with the numeric annotation), while the red curve (the mode) shows that the identity of the selected GT cycles across $\mathrm{GT}1$--$\mathrm{GT}6$ over time. Hence, the policy achieves two desirable properties simultaneously: (i) {intra-slot consensus}, which facilitates aligning the RIS beam and transmission resources to one GT for high instantaneous throughput; and (ii) {inter-slot balance}, which distributes service opportunities evenly across GTs under the TDMA constraint and supports long-term fairness.
\item Subfigures~\ref{fig:ma1} and~\ref{fig:ma2} show the MA indices for UAV~1 and UAV~10 on the $3\times3$ discrete grid. The sequences are highly concentrated in a specific region of the grid (e.g., the lower–left $2\times2$ cluster) and switch locally within that region across slots. This keeps the effective UAV–RIS incident direction stable, complements the linear phase pattern observed in~\ref{fig:ris}, and reduces mechanical movement and energy cost, while still allowing fine angular adjustments as the geometry slowly evolves.
\end{enumerate}

As shown in Fig. \ref{fig:perf}(a), the CDF of throughput for Ours is shifted to the right of all baselines, meaning a larger fraction of UAVs achieve higher link rates (notably in the 10–15 Kbps range). In Fig. \ref{fig:perf}(b), the propulsion-energy efficiency CDF for Ours lies well to the left of IA2C/CommNet/DIAL (lower energy for most cases) and is competitive with the most frugal baselines (ConseNet/FPrint), which are slightly better at a few quantiles but deliver lower throughput.

\section{Conclusion}
In this work we formulated energy–aware multi-UAV trajectory, scheduling, slot-time and RIS/MA configuration in emergency communication scenario as a networked cooperative MDP and solved it with a decentralized spatiotemporal A2C that communicates differentiable messages (state, policy fingerprints, and beliefs). Our analysis established how information and gradients propagate across the communication graph with one-hop-per-slot latency, and used a spatial discount to keep targets dominated by local interactions. Extensive experiments showed that the proposed method learns faster, attains higher long-run rewards, yields lower TD/advantage errors, and achieves a superior throughput–energy trade-off compared with IA2C, ConseNet, FPrint, DIAL and CommNet.

\bibliographystyle{Bibliography/IEEEtranTIE}
\bibliography{Bibliography/IEEEabrv,Bibliography/mybibfile.bib}\ 

@ARTICLE{10278220,
  author={Zhu, Lipeng and Ma, Wenyan and Zhang, Rui},
  journal={IEEE Communications Letters}, 
  title={Movable-Antenna Array Enhanced Beamforming: Achieving Full Array Gain With Null Steering}, 
  year={2023},
  volume={27},
  number={12},
  pages={3340-3344}}

@ARTICLE{10906511,
  author={Zhu, Lipeng and Ma, Wenyan and Mei, Weidong and Zeng, Yong and Wu, Qingqing and Ning, Boyu and Xiao, Zhenyu and Shao, Xiaodan and Zhang, Jun and Zhang, Rui},
  journal={IEEE Communications Surveys and Tutorials}, 
  title={A Tutorial on Movable Antennas for Wireless Networks}, 
  year={2025},
  volume={},
  number={},
  pages={1-1}}

@ARTICLE{10643473,
  author={Ma, Wenyan and Zhu, Lipeng and Zhang, Rui},
  journal={IEEE Transactions on Wireless Communications}, 
  title={Movable Antenna Enhanced Wireless Sensing via Antenna Position Optimization}, 
  year={2024},
  volume={23},
  number={11},
  pages={16575-16589}}

@article{chen2025space,
  title={Space-air-ground integrated network (SAGIN) in disaster management: A Survey},
  author={Chen, Ke and Zhang, Li and Zhong, Jihai},
  journal={IEEE Transactions on Network and Service Management},
  year={2025},
  publisher={IEEE}
}

@article{shi2025dynamic,
  title={Dynamic Offloading Strategy in SAGIN-based Emergency VEC: A Multi-UAV Clustering and Collaborative Computing Approach},
  author={Shi, Zhenzheng and Wang, Liang and Lin, Yaguang and Cai, Anna and Fan, Jiamin and Liu, Cong},
  journal={Vehicular Communications},
  pages={100952},
  year={2025},
  publisher={Elsevier}
}

@article{betalo2025generative,
  title={Generative AI-Driven Multi-Agent DRL for Task Allocation in UAV-Assisted EMPD within 6G-Enabled SAGIN Networks},
  author={Betalo, Mesfin Leranso and Ullah, Inam and Tesema, Fiseha Berhanu and Wu, Zongze and Li, Jianqiang and Bai, Xiaoshan},
  journal={IEEE Internet of Things Journal},
  year={2025},
  publisher={IEEE}
}

@inproceedings{xia2024uav,
  title={UAV Access Control for Urban Emergency Communications in Space-Air-Ground Integrated Networks},
  author={Xia, Guiyang and Xu, Hao and Zhou, Xiaobo and Hua, Meng and Wang, Jiangzhou},
  booktitle={2024 16th International Conference on Wireless Communications and Signal Processing (WCSP)},
  pages={235--241},
  year={2024},
  organization={IEEE}
}

@ARTICLE{10925529,
  author={Song, Fei and Wang, Zhe and Li, Jun and Shi, Long and Chen, Wen and Jin, Shi},
  journal={IEEE Transactions on Wireless Communications}, 
  title={Dynamic Trajectory and Power Control in Ultra-Dense AAV Networks: A Mean-Field Reinforcement Learning Approach}, 
  year={2025},
  volume={24},
  number={7},
  pages={5620-5634}}

@ARTICLE{11082461,
  author={Wang, Honghao and Wu, Qingqing and Gao, Ying and Chen, Wen and Mei, Weidong and Hu, Guojie and Xu, Lexi},
  journal={IEEE Transactions on Wireless Communications}, 
  title={Throughput Maximization for Movable Antenna Systems with Movement Delay Consideration}, 
  year={2025},
  volume={},
  number={},
  pages={1-1}}

@ARTICLE{10972180,
  author={Li, Zhendong and Ba, Jianle and Su, Zhou and Huang, Jinyuan and Peng, Haixia and Chen, Wen and Du, Linkang and Luan, Tom H.},
  journal={IEEE Wireless Communications}, 
  title={Movable Antennas Enabled ISAC Systems: Fundamentals, Opportunities, and Future Directions}, 
  year={2025},
  volume={},
  number={},
  pages={1-8}}

@ARTICLE{10964638,
  author={Liu, Yixin and Liang, Shaoling and Wang, Kunlun and Chen, Wen and Li, Yonghui and Karagiannidis, George K.},
  journal={IEEE Transactions on Vehicular Technology}, 
  title={Distributed Massive MIMO-Aided Task Offloading in Satellite-Terrestrial Integrated Multi-Tier VEC Networks}, 
  year={2025},
  volume={},
  number={},
  pages={1-6}}

@ARTICLE{8811733,
  author={Wu, Qingqing and Zhang, Rui},
  journal={IEEE Transactions on Wireless Communications}, 
  title={Intelligent Reflecting Surface Enhanced Wireless Network via Joint Active and Passive Beamforming}, 
  year={2019},
  volume={18},
  number={11},
  pages={5394-5409}}

@inproceedings{zhang2018fully,
  title={Fully decentralized multi-agent reinforcement learning with networked agents},
  author={Zhang, Kaiqing and Yang, Zhuoran and Liu, Han and Zhang, Tong and Basar, Tamer},
  booktitle={International conference on machine learning},
  pages={5872--5881},
  year={2018},
  organization={PMLR}
}

@article{singh2018learning,
  title={Learning when to communicate at scale in multiagent cooperative and competitive tasks},
  author={Singh, Amanpreet and Jain, Tushar and Sukhbaatar, Sainbayar},
  journal={arXiv preprint arXiv:1812.09755},
  year={2018}
}

@inproceedings{wang2023joint,
  title={Joint Phase Shift and Deployment Optimization for Multi-RIS-aided Emergency Communications},
  author={Wang, Huanwen and Lin, Lan and Xu, Wenjun},
  booktitle={2023 9th International Conference on Computer and Communications (ICCC)},
  pages={518--523},
  year={2023},
  organization={IEEE}
}

@article{mei20223d,
  title={3D-trajectory and phase-shift design for RIS-assisted UAV systems using deep reinforcement learning},
  author={Mei, Haibo and Yang, Kun and Liu, Qiang and Wang, Kezhi},
  journal={IEEE Transactions on Vehicular Technology},
  volume={71},
  number={3},
  pages={3020--3029},
  year={2022},
  publisher={IEEE}
}

@article{matracia2023comparing,
  title={Comparing aerial-RIS-and aerial-base-station-aided post-disaster cellular networks},
  author={Matracia, Maurilio and Kishk, Mustafa A and Alouini, Mohamed-Slim},
  journal={IEEE Open Journal of Vehicular Technology},
  volume={4},
  pages={782--795},
  year={2023},
  publisher={IEEE}
}

@article{foerster2016learning,
  title={Learning to communicate with deep multi-agent reinforcement learning},
  author={Foerster, Jakob and Assael, Ioannis Alexandros and De Freitas, Nando and Whiteson, Shimon},
  journal={Advances in neural information processing systems},
  volume={29},
  year={2016}
}

@inproceedings{foerster2017stabilising,
  title={Stabilising experience replay for deep multi-agent reinforcement learning},
  author={Foerster, Jakob and Nardelli, Nantas and Farquhar, Gregory and Afouras, Triantafyllos and Torr, Philip HS and Kohli, Pushmeet and Whiteson, Shimon},
  booktitle={International conference on machine learning},
  pages={1146--1155},
  year={2017},
  organization={PMLR}
}

@article{epochsupplemental,
  title={Supplemental: Deep Decentralized Multi-task Multi-agent RL under Partial Observability},
  author={Epoch, Training},
  journal={arXiv preprint arXiv:1703.06182},
  year={2017}
}

@inproceedings{gupta2017cooperative,
  title={Cooperative multi-agent control using deep reinforcement learning},
  author={Gupta, Jayesh K and Egorov, Maxim and Kochenderfer, Mykel},
  booktitle={International conference on autonomous agents and multiagent systems},
  pages={66--83},
  year={2017},
  organization={Springer}
}

@article{lowe2017multi,
  title={Multi-agent actor-critic for mixed cooperative-competitive environments},
  author={Lowe, Ryan and Wu, Yi I and Tamar, Aviv and Harb, Jean and Pieter Abbeel, OpenAI and Mordatch, Igor},
  journal={Advances in neural information processing systems},
  volume={30},
  year={2017}
}

@inproceedings{yang2018mean,
  title={Mean field multi-agent reinforcement learning},
  author={Yang, Yaodong and Luo, Rui and Li, Minne and Zhou, Ming and Zhang, Weinan and Wang, Jun},
  booktitle={International conference on machine learning},
  pages={5571--5580},
  year={2018},
  organization={PMLR}
}

@article{sukhbaatar2016learning,
  title={Learning multiagent communication with backpropagation},
  author={Sukhbaatar, Sainbayar and Fergus, Rob and others},
  journal={Advances in neural information processing systems},
  volume={29},
  year={2016}
}

@article{peng2017multiagent,
  title={Multiagent bidirectionally-coordinated nets: Emergence of human-level coordination in learning to play starcraft combat games},
  author={Peng, Peng and Wen, Ying and Yang, Yaodong and Yuan, Quan and Tang, Zhenkun and Long, Haitao and Wang, Jun},
  journal={arXiv preprint arXiv:1703.10069},
  year={2017}
}

@article{jiang2018learning,
  title={Learning attentional communication for multi-agent cooperation},
  author={Jiang, Jiechuan and Lu, Zongqing},
  journal={Advances in neural information processing systems},
  volume={31},
  year={2018}
}


\section*{Appendix A: Proof of Proposition 1}
\begin{proof}

{[Temporal domain]}:  Each UAV $j$ uses as input an augmented observation $\tilde{\mathcal{S}}_j[n]$, constructed by aggregating its own state and neighbor messages $\{\mathcal{M}_{ij}[n]\}_{i \in \mathcal{N}_j}$ recursively (Eq. (\ref{eq:recur})).
\begin{figure*}[htb]
\begin{equation}
\begin{aligned}
\tilde{\mathcal{S}}_j[n] \;&\supset \mathcal{S}_j[n] \cup  \Big\{\mathcal{S}_i[n]  \cup  \{\mathcal{M}_{ki}[n-1]\}_{k \in \mathcal{N}_{i}} \Big\}_{i \in \mathcal{N}_{j}} \\
&\supset \{\mathcal{S}_j[n]\}_{j \in \mathcal{V}_{j}}\cup \Big\{ \mathcal{S}_j[n-1] \cup \{\mathcal{M}_{ki}[n-2]\}_{k \in \mathcal{N}_{i}} \Big\}_{i \in \mathcal{V}: d_{ij}=2} \\
&\supset \{\mathcal{S}_j[n]\,\}_{i \in \mathcal{V}_{j}} \cup \{\mathcal{S}_j[n-1]\}_{i \in \mathcal{V}: d_{ij}=2} \cup \Big\{\mathcal{S}_j[n-2] \cup \{\mathcal{M}_{ki}[n-3]\}_{k \in \mathcal{N}_{i}} \Big\}_{i \in \mathcal{V}: d_{ij}=3} \\
&\quad \vdots \\
&\supset \mathcal{S}_j[n] \cup \Big\{ \mathcal{S}_j\big[n + 1 - d_{ij}\big] \Big\}_{i \in \mathcal{V} \setminus j}
\end{aligned}
 \label{eq:recur}
\end{equation}
\noindent\rule{\textwidth}{0.4pt}
\end{figure*}
This captures state information from UAVs at different hop distances, with greater delays for farther UAVs.

\noindent
{[Spatial domain]}: $\hat{\mathcal{R}}_j[n]$ at hop distance $d_{ji}$ is weighted by $\alpha^{d_{ji}}$. For $\alpha<1$, the weight decays exponentially with distance:
\begin{equation}
\alpha^{d_{ji}} \le \alpha^h, \quad \forall\, i: d_{ji} \ge h.
\end{equation}
The total contribution from neighborhoop is bounded by
\begin{equation}
\sum_{i\in \mathcal{N}_j} \alpha^{d_{ji}} \hat{\mathcal{R}}_i[n] \le |\mathcal{N}_j| \, \alpha^h \, \max_{i} |\mathcal{R}_i[n]|.
\end{equation}
Since $|\mathcal{N}_j|$ grows at most polynomially with $h$ in a sparse communication graph, while $\alpha^h$ decays exponentially, the influence of distant UAVs vanishes as $h\to\infty$. Thus, the learning target is dominated by local interactions spatiotemporally, which completes the proof.
\end{proof}

\section*{Appendix B: Proof of Proposition ~\ref{prop:neurcomm}}
\begin{proof}
The message carries previous belief and policy fingerprint (prior-decision):
$\mathcal{M}_i[n]\supset h_i[n{-}1]$, and the belief update (Eq.~\eqref{eq:neurcomm-belief}) implies
$h_j[n] \supset h_j[n{-}1]\cup \{\mathcal{S}_i[n]\}_{i\in\mathcal{V}_j}\cup\{\pi_{i}[n{-}1]\}_{i\in\mathcal{N}_j}\cup\{\mathcal{M}_i[n]\}_{i\in\mathcal{N}_j}$,
where $\mathcal{V}_j=\mathcal{N}_j\cup\{j\}$. Hence, Eq. \eqref{eq:expansion} holds, which establishes the hop-dependent delays in \eqref{eq:delayed-global}. 
\begin{figure*}[t]
\begin{equation}
\begin{aligned}
h_{j}[n] & \supset \mathcal{S}_j[n] \cup\left\{\mathcal{S}_i[n], \pi_{i} [n-1]\right\}_{i \in \mathcal{N}_{j}} \cup\left\{h_{i}[n-1]\right\}_{i \in \mathcal{V}_{j}} \\
& \supset \mathcal{S}_j[n] \cup\left\{ \mathcal{S}_i[n], \pi_{i}[n-1]\right\}_{i \in \mathcal{N}_{j}} \cup\left\{\mathcal{S}_i[n-1] \cup\left\{\mathcal{S}_k[n-1], \pi_{k}[n-2]\right\}_{k \in \mathcal{N}_{i}} \cup\left\{h_{k} [n-2]\right\}_{k \in \mathcal{V}_{i}}\right\}_{i \in \mathcal{V}_{j}} \\
& = \mathcal{S}_{j}[n-1: n] \cup\left\{\mathcal{S}_{i}[n-1: n], \pi_{i}[n-2: n-1]\right\}_{i \in \mathcal{N}_{j}} \cup\left\{\mathcal{S}_{i}[n-1], \pi_{i}[n-2]\right\}_{i \in\left\{\mathcal{V} \mid d_{j i}=2\right\}}  \cup\left\{h_{i}[n-2]\right\}_{i \in\left\{\mathcal{V} \mid d_{j i} \leq 2\right\}} \\
& \quad \cdots \\
& \quad \mathcal{S}_{j}[0: n] \cup\left\{\mathcal{S}_{i}[0: n], \pi_{i}[n-2: n-1]\right\}_{i \in \mathcal{N}_{j}} \cup\left\{\mathcal{S}_{i}[0: n-1], \pi_{i}[0: n-2]\right\}_{i \in\left\{\mathcal{V} \mid d_{j i}=2\right\}} \\
& \quad \cup \ldots \cup\left\{\mathcal{S}_{i}[0: n+1-d_{\max }], \pi_{i}[0: n-d_{\max }]\right\}_{i \in\left\{\mathcal{V} \mid d_{j i}=d_{\max }\right\}},
\end{aligned}
\label{eq:expansion}
\end{equation}
\noindent\rule{\textwidth}{0.4pt}
\end{figure*}
\end{proof}

\section*{Appendix C: Proof of Proposition 3}
\begin{proof}
Since $\mathcal{M}_i[n]\supset h_i[n{-}1]$, we obtain the expansion
\begin{equation}
\begin{aligned}
h_{j}[n] & \supset\left\{\mathcal{M}_{i}[n]\right\}_{i \in \mathcal{N}_{j}} \supset\left\{h_{i}[n-1]\right\}_{i \in \mathcal{N}_{j}} \\
& \supset\left\{\mathcal{M}_{i}[n-1]\right\}_{i \in\left\{\mathcal{V} \mid d_{j i}=2\right\}} \supset \cdots \\
& \supset\left\{\mathcal{M}_{i}[n+1-d]\right\}_{i \in\left\{\mathcal{V} \mid d_{j i}=d\right\}} \supset \cdots
\end{aligned}
\end{equation}
Thus, the message $\mathcal{M}_i[\tau]$ of UAV $i$ appears on the computation graph of UAV $j$ at time $\tau+d_{ji}-1$. Equivalently, parameters $\{\nu_i,\lambda_i\}$ receive gradients from $\mathcal{L}(\theta_j)$ and $\mathcal{L}(\phi_j)$ for all $j\neq i$, except for the first $d_{ji}-1$ experience samples due to hop latency. If $d_{\max}\ll|\mathcal{B}|$, then $\{\nu_i,\lambda_i\}$ receive almost all gradients from the loss signals of other UAVs. 
\end{proof}

\end{document}